\documentclass[a4paper,11pt]{article}

\pdfoutput=1

\usepackage{amsmath,amssymb,mathtools}
\usepackage{color}
\usepackage{graphicx}
\usepackage{subfigure}
\usepackage{cite}
\usepackage[colorlinks=true,linkcolor=blue, citecolor=red, urlcolor=blue, bookmarks]{hyperref}
\usepackage{multirow,makecell} %% multirow and multicolumn
\usepackage[figuresright]{rotating} %% rotating tables
\usepackage{textcomp}
\usepackage{wasysym}
\usepackage{diagbox}
\usepackage{ulem}

\usepackage[utf8]{inputenc}
\usepackage[T1]{fontenc}

\newcommand{\mc}[3]{\multicolumn{#1}{#2}{#3}}

\newcommand{\red}[1]{\textcolor[rgb]{1.00,0.00,0.00}{#1}}
\newcommand{\green}[1]{\textcolor[rgb]{0.00,1.00,0.00}{#1}}
\newcommand{\blue}[1]{\textcolor[rgb]{0.00,0.00,1.00}{#1}}

\usepackage[text={17.1cm,24.6cm},centering]{geometry} %% thanks to Chao Wu

\numberwithin{equation}{section}

\def \be {\begin{equation}}
\def \ee {\end{equation}}
\def \ba {\begin{array}}
\def \ea {\end{array}}
\def \bea{\begin{eqnarray}}
\def \eea{\end{eqnarray}}
\def \nn {\nonumber}

\def \a {\alpha}
\def \b {\beta}
\def \g {\gamma}
\def \c {\gamma}
\def \G {\Gamma}
\def \d {\delta}
\def \D {\Delta}
\def \e {\epsilon}
\def \ve {\varepsilon}
\def \m {\mu}
\def \n {\nu}
\def \k {\kappa}

\def \lam {\lambda}

\def \s {\sigma}
\def \S {\Sigma}
\def \r {\rho}
\def \o {\omega}
\def \O {\Omega}
\def \th {\theta}
\def \vth {\vartheta}
\def \Th {\Theta}
\def \ph {\phi}
\def \vph {\varphi}
\def \vphi {\varphi}
\def \t {\tau}
\def \z {\zeta}

\def \bA {\mathbf A}
\def \bB {\mathbf B}
\def \bC {\mathbf C}
\def \bD {\mathbf D}
\def \bE {\mathbf E}
\def \bF {\mathbf F}
\def \bG {\mathbf G}
\def \bH {\mathbf H}
\def \bI {\mathbf I}
\def \bJ {\mathbf J}
\def \bK {\mathbf K}
\def \bL {\mathbf L}
\def \bM {\mathbf M}
\def \bN {\mathbf N}
\def \bO {\mathbf O}
\def \bP {\mathbf P}
\def \bQ {\mathbf Q}
\def \bR {\mathbf R}
\def \bS {\mathbf S}
\def \bT {\mathbf T}
\def \bU {\mathbf U}
\def \bV {\mathbf V}
\def \bW {\mathbf W}
\def \bX {\mathbf X}
\def \bY {\mathbf Y}
\def \bZ {\mathbf Z}

\def \cA {\mathcal A}
\def \cB {\mathcal B}
\def \cC {\mathcal C}
\def \cD {\mathcal D}
\def \cE {\mathcal E}
\def \cF {\mathcal F}
\def \cG {\mathcal G}
\def \cH {\mathcal H}
\def \cI {\mathcal I}
\def \cJ {\mathcal J}
\def \cK {\mathcal K}
\def \cL {\mathcal L}
\def \cM {\mathcal M}
\def \cN {\mathcal N}
\def \cO {\mathcal O}
\def \cP {\mathcal P}
\def \cQ {\mathcal Q}
\def \cR {\mathcal R}
\def \cS {\mathcal S}
\def \cT {\mathcal T}
\def \cU {\mathcal U}
\def \cV {\mathcal V}
\def \cW {\mathcal W}
\def \cX {\mathcal X}
\def \cY {\mathcal Y}
\def \cZ {\mathcal Z}

\def \rA {\mathrm A}
\def \rB {\mathrm B}
\def \rC {\mathrm C}
\def \rD {\mathrm D}
\def \rE {\mathrm E}
\def \rF {\mathrm F}
\def \rG {\mathrm G}
\def \rH {\mathrm H}
\def \rI {\mathrm I}
\def \rJ {\mathrm J}
\def \rK {\mathrm K}
\def \rL {\mathrm L}
\def \rM {\mathrm M}
\def \rN {\mathrm N}
\def \rO {\mathrm O}
\def \rP {\mathrm P}
\def \rQ {\mathrm Q}
\def \rR {\mathrm R}
\def \rS {\mathrm S}
\def \rT {\mathrm T}
\def \rU {\mathrm U}
\def \rV {\mathrm V}
\def \rW {\mathrm W}
\def \rX {\mathrm X}
\def \rY {\mathrm Y}
\def \rZ {\mathrm Z}

\def \mA {\mathcal A}
\def \mB {\mathcal B}
\def \mC {\mathcal C}
\def \mD {\mathcal D}
\def \mE {\mathcal E}
\def \mF {\mathcal F}
\def \mG {\mathcal G}
\def \mH {\mathcal H}
\def \mI {\mathcal I}
\def \mJ {\mathcal J}
\def \mK {\mathcal K}
\def \mL {\mathcal L}
\def \mM {\mathcal M}
\def \mN {\mathcal N}
\def \mO {\mathcal O}
\def \mP {\mathcal P}
\def \mQ {\mathcal Q}
\def \mR {\mathcal R}
\def \mS {\mathcal S}
\def \mT {\mathcal T}
\def \mU {\mathcal U}
\def \mV {\mathcal V}
\def \mW {\mathcal W}
\def \mX {\mathcal X}
\def \mY {\mathcal Y}
\def \mZ {\mathcal Z}

\def \p {\partial}

\def \f {\frac}
\def \df {\dfrac}
\def \na {\nabla}
\def \scl {\ell}
\def \mb {\mathbf}
\def \ma {\mathcal}
\def \mf {\mathfrak}

\def \lt {\left}
\def \rt {\right}

\def \sl {\slashed}
\def \la {\leftarrow}
\def \ra {\rightarrow}

\def \da {\downarrow}
\def \lra {\leftrightarrow}
\def \sr {\sqrt}
\def \td {\tilde}
\def \hs {\hspace}
\def \pp {\propto}
\def \inf {\infty}
\def \tph {{\tilde\phi}}
\def \tps {{\tilde\psi}}
\def \hph {{\hat\phi}}
\def \hps {{\hat\psi}}
\def \ho {{\hat\omega}}

\def \lag {\langle}
\def \rag {\rangle}
\def \ph  {\phantom}
\def \ul  {\underline}
\def \cas {\circledast}
\def \cc  {\circ}
\def \ccc {\circledcirc}

\def \app  {\approx}
\def \dd {\!\!\mathrm{d}}
\def \ep {\mathrm{e}}
\def \ii {\mathrm{i}}

\def \hatt {{\hat t}}
\def \hatr {{\hat r}}

\def \arcsinh {\mathop{\rm arcsinh}}
\def \arccosh {\mathop{\rm arccosh}}
\def \arctanh {\mathop{\rm arctanh}}

\def \Re {\mathop{\textrm{Re}}}
\def \Im {\mathop{\textrm{Im}}}
\def \tr {\textrm{tr}}
\def \Tr {{\mathop\textrm{Tr}}}
\def \Str {\mathop{\textrm{Str}}}
\def \STr {\mathop{\textrm{STr}}}
\def \diag {\mathop{\textrm{diag}}}
\def \Sdet {\mathop{\textrm{Sdet}}}
\def \SDet {\mathop{\textrm{SDet}}}

\def \and {{~\textrm{and}~}}
\def \with {{~\textrm{with}~}}

\newcommand{\bra}[1]{\ensuremath{\left\langle#1\right|}}
\newcommand{\ket}[1]{\ensuremath{\left|#1\right\rangle}}

\begin{document}
\sloppy

\newpage
\setcounter{page}{1}

\begin{center}{\Large \textbf{
Efficient Representation of Gaussian Fermionic Pure States in Non-Computational Bases\\
}}\end{center}

% TODO: write the author list here. Use first name (+ other initials) + surname format.
% Separate subsequent authors by a comma, omit comma and use "and" for the last author.
% Mark the corresponding author with a superscript star.
\begin{center}
Babak Tarighi\textsuperscript{1},
Reyhaneh Khasseh\textsuperscript{2} and
M. A. Rajabpour\textsuperscript{1$\star$}
\end{center}

\begin{center}
{\bf 1} Universidade Federal Fluminense, Niter\'oi, Brazil
\\
{\bf 2} Theoretical Physics III, Center for Electronic Correlations and Magnetism, Institute of Physics, University of Augsburg, D-86135 Augsburg, Germany
%\\
%${}^\star$ {\small \sf arashjafarizadeh@id.uff.br}
\end{center}

\begin{center}
\today
\end{center}

% For convenience during refereeing (optional),
% you can turn on line numbers by uncommenting the next line:
%\linenumbers

\section*{Abstract}
This paper introduces an innovative approach for representing Gaussian fermionic states, pivotal in quantum spin systems and fermionic models, within a range of alternative quantum bases. We focus on transitioning these states from the conventional computational \(\sigma^z\) basis to more complex bases, such as \((\phi, \frac{\pi}{2}, \alpha)\), which are essential for accurately calculating critical quantities like formation probabilities and Shannon entropy. Our methodology can be advantageous in quantum quench studies and quantum tomography, where alternative basis representations significantly enhance optimization processes.  We present a novel algorithm that not only simplifies the basis transformation but also reduces computational complexity, making it feasible to calculate amplitudes of large systems efficiently. Our key contribution is a technique that translates amplitude calculations into the Pfaffian computation of submatrices from an antisymmetric matrix, a process facilitated by understanding domain wall relationships across different bases. 
As an application, we will determine the formation probabilities for various bases and configurations within the critical transverse field Ising chain, considering both periodic and open boundary conditions. We aim to categorize the configurations and bases by examining the universal constant term that characterizes the scaling of the logarithm of the formation probability in the periodic system, as well as the coefficient of the logarithmic term in the case of open systems. In the open system scenario, this coefficient is influenced by the central charge and the conformal weight of the boundary condition-changing operator. This work is set to expand the toolkit available for researchers in quantum information theory and many-body physics, providing a more efficient and elegant solution for exploring Gaussian fermionic states in non-standard quantum bases.

\vspace{10pt}
\noindent\rule{\textwidth}{1pt}
%\tableofcontents\thispagestyle{fancy}
\tableofcontents
\noindent\rule{\textwidth}{1pt}
\vspace{10pt}

\section{Introduction}\label{sec:Introduction}

Gaussian fermionic states play a pivotal role as fundamental components in the realm of spin systems, serving as precise eigenstates for systems directly mappable to quadratic free fermionic models like the XY spin chain \cite{LIEB1961407,Peschel2001,Zanardi2008}. They are not only integral in such direct mappings but also stand as highly effective approximations for the ground states of interactive systems\cite{Thouless1960,Kraus2010,Terhal2023,Bravyi2017}. They are also the bases for the simulation of free fermionic
computation \cite{Bravyi2012}. Their significance is further highlighted as they often represent the sole states amenable to analytical scrutiny or as foundational elements in constructing more complex states for probing interacting Hamiltonians.

In their conventional form, Gaussian states are typically expressed in a manner that aligns seamlessly with the computational basis (the $\sigma^z$ basis), a choice stemming from their inherent definition. Yet, in a multitude of practical scenarios, there arises a necessity to comprehend and articulate these states in alternate bases. For instance, in scenarios where the crux of the physics under study is better represented in the $\sigma^x$ basis, it becomes imperative to reframe these states within this new basis to accurately compute pertinent quantities like global entanglement \cite{Wei2005} formation probabilities \cite{Essler1994,Essler1995,Shiroishi2001,Franchini2005,Stephan2013,NR2016,Rajabpour2020} or Shannon entropy\cite{Stephan2009,Stephan2010,Alcaraz2013,Alcaraz2014,Tarighi2022,Jiaju2023}.

Another scenario where such a basis shift might be useful is in the study of quantum quenches. Here, the initial state, being Gaussian, needs to be evolved under a Hamiltonian that exhibits a markedly simpler structure on a different basis. Similarly, in the realm of quantum tomography, leveraging Gaussian states for optimization processes could be substantially more efficient when approached from an alternative basis perspective \cite{Najafi2021, Tarighi2024}. This need for a basis shift underscores the importance of devising an efficient methodology for expressing Gaussian fermionic states beyond the traditional computational framework, thereby broadening their applicability and enhancing our understanding of various quantum phenomena.

Consider an arbitrary qubit in the computational basis, to change the basis to a new generic basis, one can use the following unitary matrix
\begin{equation}{\label{Generic U matrix}}
    \bold{U}_{(\phi,\theta,\alpha)}=\left(
\begin{array}{ccccccc}
    \cos{\frac{\theta}{2}}    & \sin{\frac{\theta}{2}}e^{-i\alpha}  \\
 \sin{\frac{\theta}{2}}e^{-i\phi}  &  -\cos{\frac{\theta}{2}}e^{-i(\alpha+\phi)}   \\
\end{array}
\right).
\end{equation}
For example, $(\phi,\theta,\alpha)=(0,\frac{\pi}{2},0)$ and $(\frac{\pi}{2},\frac{\pi}{2},0)$ are the $\sigma^x$ and $\sigma^y$ bases respectively.

To write a state expressed in computational basis in the $(\phi,\theta,\alpha)$ basis one may  use the mapping

\begin{equation}{\label{U matrix base change}}
    \ket{\psi}_{(\phi,\theta,\alpha)} = \bold{U}_{(\phi,\theta,\alpha)}\otimes \bold{U}_{(\phi,\theta,\alpha)}...\otimes \bold{U}_{(\phi,\theta,\alpha)} \ket{\psi}_z.
\end{equation}
However, this is neither efficient nor elegant because it requires first writing the state in the $\sigma^z$ basis and then multiple matrix multiplications with the complexity growing exponentially. This problem is most pronounced if one is interested in the amplitude of a particular configuration in the $(\phi,\theta,\alpha)$ basis. In our paper, we present a highly exact efficient technique for representing any arbitrary Gaussian pure state in the $(\phi, \frac{\pi}{2}, \alpha)$ basis. This method introduces an algorithm whose complexity increases polynomially, allowing for the precise calculation of the amplitudes. Significantly, we have translated the task of determining the amplitude in the specified basis into the computation of the Pfaffian of a submatrix derived from a specific antisymmetric matrix. This achievement was accomplished by leveraging the relationship between the domain walls in the mentioned basis and the $\sigma^z$ basis. Note that our formulas are independent of the system's dimension or the range of the underlying couplings in the system.

To demonstrate the effectiveness of our approach, we apply our formulas to compute the formation probabilities of various bit strings in the ground state of the critical transverse field Ising chain across different bases. In the case of periodic chains, these probabilities have been extensively analyzed in the $\sigma^z$ basis, as detailed in \cite{Rajabpour2020}. This analysis revealed that the logarithm of the probabilities exhibits a linear relationship with the system size, accompanied by a subleading constant term linked to the boundary entropy introduced in \cite{AL1991}. This constant term indicates the conformal boundary condition towards which each configuration flows. We extend the findings of \cite{Rajabpour2020} to include arbitrary bases in the $xy$ plane. By evaluating these probabilities for large system sizes ($L\approx1000$), we identify the constant terms and, consequently, the conformal boundary condition each configuration approaches. Additionally, we investigate these probabilities for critical open chains. In such instances, boundary conformal field theory predicts that the probability's logarithm should exhibit a leading linear term plus a secondary logarithmic term, as suggested in \cite{Peschel1988}, see also \cite{Stephan2013}. The coefficient of this logarithmic term is influenced by the central charge and the potential conformal weight of the boundary condition-changing operator. We calculate this coefficient for various configurations and categorize them across different bases.

The structure of the paper is outlined as follows: Section (\ref{sec:GPS}) begins with an introduction to Gaussian pure states, followed by a demonstration of how to determine the amplitudes in the computational basis via the Pfaffian of the submatrices of the matrix defining the Gaussian state. Section (\ref{sec:GPSsigmax}) then presents a precise formula for obtaining the amplitudes of the Gaussian state in the $\sigma^x$ basis. This formula enables the effortless calculation of specific configuration amplitudes in this basis for large sizes $L>1000$. Moving forward, in Section (\ref{sec:phi basis}), we broaden these findings to the $(\phi,\frac{\pi}{2},0)$ basis, utilizing a method akin to the previous sections. In Section (\ref{sec:phi-alpha basis}), we introduce a formula to ascertain the Gaussian pure state in the $(\phi,\frac{\pi}{2},\alpha)$ basis. 
In section (\ref{sec:probabilities}), we initially present generic formulas for computing the probabilities of various configurations across different bases for a Gaussian state. Following that, we apply these formulas to evaluate the formation probabilities of diverse bit strings in the ground state of both open and periodic critical transverse field Ising chains, aiming to identify the universal quantities.

The paper culminates in the final section with our conclusions. Additionally, three appendices accompany the article, providing numerous examples to further elucidate the presented concepts.

\section{Gaussian pure states}\label{sec:GPS}

In this section, we first introduce the Gaussian pure states and list a few important properties that are directly related to this work. Consider the following Gaussian pure state 
\begin{equation}{\label{GPS0}}
    \ket{\bold{R},0} = \frac{1}{\mathcal{N}_R}e^{{\frac{1}{2}\sum_{i,j}^Lc^{\dagger}_{i}r_{ij}c^{\dagger}_{j}}}\ket{0},
\end{equation}
where $\mathcal{N}_R=\det(\bold{I}+\bold{R}^{\dagger}.\bold{R})^{\frac{1}{4}}$ and $c_j\ket{0}=0, \forall j$. Note that without losing generality, we can always consider the matrix $\bold{R}$  anti-symmetric. We will occasionally also use the notation $\bold{c}^{\dagger}.\bold{R}.\bold{c}^{\dagger}=\sum_{i,j}^Lc^{\dagger}_{i}r_{ij}c^{\dagger}_{j}$.
To write the above state in the configuration basis of the bit strings, one can write the state in the fermionic coherent basis and then put the Grassmann variables of sites where there is no fermion zero and Berezin integrate over the sites where there is a fermion\cite{NR2016}. This procedure leads to the following elegant formula:
\begin{equation}{\label{GPS0-pfaffinho}}
    \ket{\bold{R},0} = \frac{1}{\mathcal{N}_R}\sum_{\mathcal{I}}\bold{pf}\ R_{\mathcal{I}}{\ket{\mathcal{I}}},
\end{equation}
where $\mathcal{I}$ is the bit string configuration,  $\bold{R}_{\mathcal{I}}$ is the submatrix of the matrix $\bold{R}$ in which we removed the rows and columns corresponding to the sites that there is no fermion. Finally, $\bold{pf}$ indicates the pfaffian of the matrix. From now on we will call the $\bold{pf}\ R_{\mathcal{I}}$ the {\it{pfaffinho}} of the the matrix $\bold{R}$. The maximum number of non-zero pfaffinhos for each anti-symmetric matrix is $2^{L-1}$. The implicit version of the formula mentioned has already been presented in reference \cite{Bravyi2005}, and the explicit form can be found in \cite{Becca-Sorella2017}.

For an example, consider the following  state with $L=4$:

\begin{eqnarray}{\label{GPS-exampleo}}
    \ket{\bold{R},0000} &=& \frac{1}{\mathcal{N}_R}(\ket{0000} +     
    r_{12}\ket{1100} + r_{13}\ket{1010} +\nonumber\\
    && r_{14}\ket{1001} +  r_{23}\ket{0110} + r_{24}\ket{0101} + \nonumber\\
    && r_{34}\ket{0011} + 
    \bold{pf} \bold{R}  \ket{1111}).
\end{eqnarray}
Now consider the following mapping called Jordan-Wigner (JW) transformation:

\begin{equation}{\label{JW}}
\begin{aligned}
    c_l = \prod_{j<l}(-\sigma^{z}_j)\sigma^{-}_l, 
    \\
    c^{\dagger}_l = \prod_{j<l}(-\sigma^{z}_j)\sigma^{+}_l.
\end{aligned}
\end{equation}
It is easy to show that $\sigma_l^z=2c_l^{\dagger}c_l-1$ which means that one can write the above state in the $\sigma^z$ basis by just the substitutions $\ket{0}\rightarrow \ket{\downarrow}_z$ and $\ket{1}\rightarrow \ket{\uparrow}_z$.

   The general definition of a Gaussian pure state is
 
\begin{equation}{\label{GPS-general}}
    \ket{\bold{R},\mathcal{C}} = \frac{1}{\mathcal{N}_R}e^{{\frac{1}{2}\sum_{i,j}^La_{i}r_{ij}a_{j}}}\ket{\mathcal{C}},
\end{equation}
where now the $a_j=c_j(c_j^{\dagger})$ if there is (no) fermion at site $j$ of the configuration $\mathcal{C}$. In the Appendix \ref{sec:GaussianStateGeneral}, we show that the most general Gaussian state can be transformed into (\ref{GPS-general}).  Note that the equation (\ref{GPS-general}) may show a vanishing
contribution of the standard vacuum $\ket{0}$. To write the above state in the configuration basis it is more convenient we first define ${\ket{\mathcal{C}}}={\ket{n_1,n_2,...,n_L}}$ and ${\ket{\mathcal{I}}}={\ket{m_1,m_2,...,m_L}}$, where $n_j,m_j\in\{0,1\}$.
Then following the same procedure as above, one can write 

\begin{equation}{\label{GPS-general-pfaffinho}}
    \ket{\bold{R},\mathcal{C}} = \frac{1}{\mathcal{N}_R}\sum_{\mathcal{I}}\text{sgn}(\mathcal{C},\mathcal{I})\bold{pf}\ R_{\mathcal{I}}(\mathcal{C}){\ket{\mathcal{I}}},
\end{equation}
where now $R_{\mathcal{I}}(\mathcal{C})$ is the submatrix of the matrix $\bold{R}$ defined such that we keep the rows and columns $j\in\{1,2,...,L\}$ such that $|n_j-m_j|=1$. The sign can be found using the following equation:

\begin{equation}{\label{GPS-general-sign}}
 \text{sgn}(\mathcal{C},\mathcal{I})=\prod_{i=2}^L(-1)^{|n_i-m_i|\sum_{j<i}n_j}.   
\end{equation}
For an example, consider the following  state with $L=4$:

\begin{eqnarray}{\label{GPS-exampleo}}
    \ket{\bold{R},1010}&=&\frac{1}{\mathcal{N}_R}(\ket{1010}-r_{34}\ket{1001} +r_{14}\ket{0011} \nonumber\\    
     &+&r_{23}\ket{1100}-r_{12} \ket{0110} -r_{13} \ket{0000} \nonumber\\
     &-&r_{24}\ket{1111} +\bold{pf} \bold{R}  \ket{0101}     ).
\end{eqnarray}

The Gaussian pure state defined as (\ref{GPS-general-pfaffinho}) has the following remarkable property: consider the state $\ket{\bold{R},\mathcal{C}}$, then as far as the configuration $\ket{\mathcal{C}'}$ has non-zero amplitude one can always find an $\bold{R}'$ matrix such that:

\begin{equation}{\label{R matrix changes}}
\ket{\bold{R},\mathcal{C}} =\ket{\bold{R}',\mathcal{C}'}.  
\end{equation}
To get the matrix $\bold{R}'$ from $\bold{R}$ and the configurations ${\ket{\mathcal{C}}}={\ket{n_1,n_2,...,n_L}}$ and ${\ket{\mathcal{C}'}}={\ket{n'_1,n'_2,...,n'_L}}$ one can do the following: First consider the set of configurations ${\ket{\mathcal{I}'}}={\ket{m'_1,m'_2,...,m'_L}}$ that can be obtained from ${\ket{\mathcal{C}'}}$ by flipping just two spins. Then the elements of the matrix $\bold{R}'$, i.e. $r'_{ij}$, can be obtained as follows:

\begin{equation}{\label{R' matrix elements}}
 r'_{ij}=\text{sgn}(\mathcal{C},\mathcal{C}')\frac{\text{sgn} (\mathcal{C},\mathcal{I}')}{\text{sgn}(\mathcal{C}',\mathcal{I}')} \frac{\bold{pf}\ R_{\mathcal{CI'}}}{\bold{pf}\ R_{\mathcal{CC'}}},
\end{equation}
where the sets

\begin{eqnarray}{\label{the sets}}
  \mathcal{CI'}&=&\{\forall j|n_j-m'_j\neq0\},\\
  \mathcal{CC'}&=&\{\forall j|n_j-n'_j\neq0\},
\end{eqnarray}
and the signs $\text{sgn}(\mathcal{C},\mathcal{I}')$ and $\text{sgn}(\mathcal{C},\mathcal{C}')$ are defined as (\ref{GPS-general-sign}).

For an example, consider the  state $\ket{\bold{R},101000}$ 
with $L=6$ we can change the base configuration to $\ket{011101}$ with the following $\bold{R}'$ matrix:

\begin{equation}{\label{R' matrix elements}}
\bold{R'}= \frac{1}{\bold{pf}\ R_{1246}}\left(
\begin{array}{ccccccc}
 0 & -r_{46} & \bold{pf}\ R_{2346} & r_{26} & -\bold{pf}\ R_{2456} & -r_{24}\\
 r_{46} & 0 & -\bold{pf}\ R_{1346} & -r_{16} & \bold{pf}\ R_{1456} & r_{14}\\
-\bold{pf} R_{2346}& \bold{pf} R_{1346}& 0 & \bold{pf}\ R_{1236} & -\bold{pf}\ R & -\bold{pf}\ R_{1234}\\
 -r_{26} & r_{16} & -\bold{pf}\ R_{1236} & 0 & -\bold{pf}\ R_{1256} & -r_{12}\\
\bold{pf}\ R_{2456} & -\bold{pf}\ R_{1456} & \bold{pf}\ R & \bold{pf}\ R_{1256} & 0 & \bold{pf}\ R_{1245}\\
r_{24} & -r_{14} & \bold{pf}\ R_{1234} & r_{12} & -\bold{pf}\ R_{1245} & 0\\
\end{array}
\right).
\end{equation}

For further details, see the Appendix \ref{sec:AppExamplesMixed}.

It is worth mentioning that the existence of the matrix $\bold{R}'$ is guaranteed because of many polynomial relations between pfaffinhos that can be obtained from the equation (\ref{R matrix changes}). Here, these equations are the consequences of the assumption that the matrix $\bold{R}'$ exists, which we left unproven.

\section{Gaussian pure states in the $\sigma^x$ basis}\label{sec:GPSsigmax}

In this section, we provide the form of the Gaussian states introduced in the previous section in the $\sigma^x$ basis, i.e. sequence of $+$ and $-$.   To write states such as  (\ref{GPS0}) in the $\sigma^x$ basis, as we mentioned in the introduction, one may first write it in the $\sigma^z$ basis as we outlined in the previous section and then use the mapping

\begin{equation}{\label{U matrix base change}}
    \ket{\psi}_x = \bold{U}\otimes \bold{U}...\otimes \bold{U} \ket{\psi}_z,
\end{equation}
where $\bold{U}$ is
\begin{equation}\label{eq: M matrix}
\begin{split}
\bold{U}= \frac{1}{\sqrt{2}}\left(
\begin{array}{ccccccc}
    1    & 1 \\
 1 &  -1  \\
\end{array}
\right).
\end{split}
\end{equation}
However, this is not efficient, especially if one is interested in the amplitude of a particular sequence for large systems. The more elegant method is to directly get the amplitudes using the $\bold{R}$ matrix. To achieve this goal, it is most convenient to work with domain walls in the $\sigma^x$ basis. We first start with the state  (\ref{GPS0}).
The remarkable thing is that there is a $\tilde{\bold{R}}$ matrix such that the state 
\begin{figure*}[htp]
\includegraphics[width=180mm]{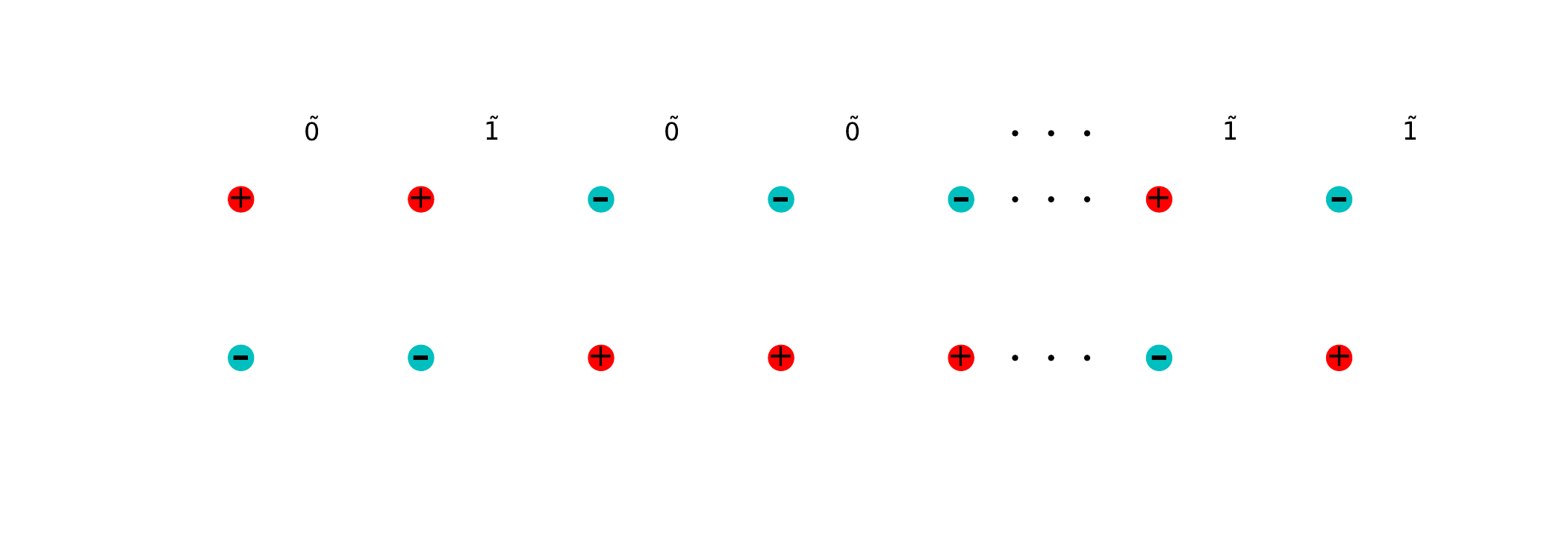}
	\caption{ The domain wall basis of the two configurations $\ket{++---....,+-}$ and $\ket{--+++....,-+}$ is $\ket{\tilde{0}\tilde{1}\tilde{0}\tilde{0},...,\tilde{1},\tilde{1}}$. Using the same logic, one can map all the configurations in the $\sigma^x$ basis to the domain wall basis. }
	\label{fig:domain_wall}
\end{figure*}
\begin{equation}{\label{Domainwall0}}
    \ket{\tilde{\bold{R}},\tilde{0}} = \frac{1}{\mathcal{N}_{\tilde{R}}}e^{{\frac{1}{2}\sum_{i,j}^Lc^{\dagger}_{i}\tilde{r}_{ij}c^{\dagger}_{j}}}\ket{\tilde{0}},
\end{equation}
describes the $\ket{\bold{R},0} $ in the domain wall basis of the $\sigma^x$ basis. In other words, consider an arbitrary sign sequence such as $\ket{++---....,+-}$ in the $\sigma^x$ basis. The domain wall form of this sequence is $\ket{\tilde{0}\tilde{1}\tilde{0}\tilde{0},...,\tilde{1},\tilde{1}}$, note that we also consider the domain wall between the last spin with the first spin too, see Figure \ref{fig:domain_wall}. Obviously, if we change the direction of all spins in the $\sigma^x$ basis, the domain wall configuration would be the same, indicating that the mapping is two to one. If we assume the matrix $\tilde{\bold{R}}$ exists and describe domain walls as we outlined, then one can write 

\begin{equation}{\label{GPS0-x basis}}
    \ket{\bold{R},0} = \frac{1}{\sqrt{2}\mathcal{N}_{\tilde{R}}}\sum_{\mathcal{S}}sgn(\mathcal{S})\bold{pf}\ \tilde{R}_{\mathcal{S}}{\ket{\mathcal{S}}}_x,
\end{equation}
where ${\ket{\mathcal{S}}}$ is a sequence of $+$ and $-$ and the $\tilde{R}_{\mathcal{S}}$ is a submatrix of the matrix $\tilde{\bold{R}}$ in which we first find the domain wall configuration of $\mathcal{S}$ and then we keep the rows and columns corresponding to the sites that there is a domain wall. The $sgn(\mathcal{S})$ is simply +1(-1) for even(odd) number of $-$ in the sequence.

%the following two operators

%\begin{eqnarray}{\label{two operators}}
% \mathcal{O}^z_j=\sigma_j^z,\hspace{2cm}\mathcal{O}^{xx}_{j,j+1}=\sigma_j^x\sigma_{j+1}^x,
%\end{eqnarray}
To find the explicit form of the matrix $\tilde{\bold{R}}$ we first notice that in  the spirit of Kramers-Wannier duality one expects the following identities:
\begin{eqnarray}{\label{KW1}}
\bra{0,\bold{R}} \sigma_j^x\sigma_{j+1}^x\ket{\bold{R},0}&=&-\bra{0,\tilde{\bold{R}}} \sigma_j^z\ket{\tilde{\bold{R}},0},\\
{\label{KW22}}
\bra{0,\tilde{\bold{R}}} \sigma_j^x\sigma_{j+1}^x\ket{\tilde{\bold{R}},0}&=&-\bra{0,\bold{R}} \sigma_{j+1}^z\ket{\bold{R},0},
\end{eqnarray}
for $j=1,2,...,L$ and $\sigma_{L+1}^x=+\sigma_{1}^x$. We start with the first equation, i.e. (\ref{KW1}). Using the JW transformation we can show that $\sigma_j^x\sigma_{j+1}^x=(c_j^{\dagger}-c_j)(c_{j+1}^{\dagger}+c_{j+1})$. We also define the following matrices:
\begin{eqnarray}{\label{C defenitions}}
C_{jk}(R)&=&\bra{0,\bold{R}} c_j^{\dagger}c_k\ket{\bold{R},0},\\
\label{ G defenitions}
G_{jk}(R)&=&\bra{0,\bold{R}} (c_j^{\dagger}-c_j)(c_k^{\dagger}+c_k)\ket{\bold{R},0}.
\end{eqnarray}
These matrices for our Gaussian states have the following forms:
\begin{eqnarray}{\label{C and G}}
\bold{C}&=&\bold{I}-\bold{Q},\\
\bold{G}&=&\bold{I}+\bold{Q}^T\cdot(\bold{R}-\bold{I})+\bold{R}^*\cdot\bold{Q}^T-\bold{Q},
\end{eqnarray}
where $\bold{Q}=(\bold{I}-\bold{R}^*\cdot\bold{R})^{-1}$.
Then the condition (\ref{KW1}) means:
\begin{eqnarray}{\label{KW1 G and C}}
G_{j,j+1}(R)&=&1-2C_{j,j}(\tilde{R}),\\
G_{L,1}(R)&=&2C_{L,L}(\tilde{R})-1,
\end{eqnarray}
where $j=1,2,...,L-1$. Following the same calculations for the equality (\ref{KW22}) one can find
\begin{eqnarray}{\label{KW2 G and C}}
G_{j,j+1}(\tilde{R})&=&1-2C_{j+1,j+1}(R),\\
G_{L,1}(\tilde{R})&=&2C_{1,1}(R)-1.
\end{eqnarray}

A solution for these equations can be found as follows:

\begin{equation}{\label{R tilda}}
    \Tilde{\bold{R}} = (\bold{I}+\bold{H}\cdot\bold{P})\cdot(\bold{H}\cdot\bold{P}-\bold{I})^{-1}.
\end{equation}
where
\begin{eqnarray}{\label{G tilda}}
\bold{H} = (\bold{R}-\bold{I})\cdot(\bold{R}+\bold{I})^{-1},
\end{eqnarray}
and
\begin{equation}\label{eq: P matrix}
\begin{split}
\bold{P}= \left(
\begin{array}{ccccccc}
 0 & 0 & 0 & 0 & \hdots & 0 & 1 \\
 -1 & 0 & 0 & 0 & \hdots & 0 & 0 \\
0 & -1 & 0 & 0 & \hdots & 0 & 0 \\
\vdots& \vdots& \vdots& \vdots& \ddots& \vdots& \vdots \\
 0 & 0 & 0 & 0 & \hdots & 0 & 0 \\
0 & 0 & 0 & 0 & \hdots & 0 & 0 \\
0 & 0 & 0 & 0 & \hdots & -1 & 0 \\
\end{array}
\right).
\end{split}
\end{equation}
%The equation (\ref{R tilda}) also provides a solution to the equation (\ref{KW22}).

In summary, to write the state (\ref{GPS0}) in the $\sigma^x$ basis, one needs to first find the matrix $\Tilde{\bold{R}}$ then for each sequence of signs, one needs to find the domain walls configurations. The amplitudes can be derived using the corresponding pfaffinhos of the matrix $\Tilde{\bold{R}}$.

An example of $L=3$ is
\begin{eqnarray}{\label{GPS-example-sigmax}}
    \ket{\bold{R},000}&=&\frac{1}{\sqrt{2}\mathcal{N}_{\tilde{R}}}
    \Big{(}\ket{+++}-\frac{b}{a}\ket{++-}-\frac{c}{a}\ket{+-+}\nonumber\\
    &+&\frac{d}{a}\ket{+--}-\frac{d}{a}\ket{-++}+\frac{c}{a}\ket{-+-}\nonumber\\
    &+&\frac{b}{a}\ket{--+} -\ket{---}    \Big{)},
\end{eqnarray}
where $\mathcal{N}_{\tilde{R}} = 2 \left(\frac{\sqrt{1 + r_{12}^2 + r_{13}^2 + r_{23}^2}}{|a|}\right)
$. To enhance clarity and readability in our discussions and formulations, from this point forward, we will represent states in the following standardized form
\begin{eqnarray}{\label{GPS-example-sigmax}}
    \ket{\bold{R},000}&=&\frac{1}{\sqrt{2}\mathcal{Z}}
    \Big{(}a\ket{+++}-b\ket{++-}-c\ket{+-+}\nonumber\\
    &+&d\ket{+--}-d\ket{-++}+c\ket{-+-}\nonumber\\&+&b\ket{--+} 
    -a\ket{---}    \Big{)},
\end{eqnarray}
where 
\begin{eqnarray}{\label{abcd}}
a&=&1+r_{12}+r_{13}+r_{23},\nonumber\\
b&=&1+r_{12}-r_{13}-r_{23},\nonumber\\
c&=&1-r_{12}+r_{13}-r_{23},\nonumber\\
d&=&1-r_{12}-r_{13}+r_{23},\nonumber
\end{eqnarray}
and $\mathcal{Z}=2\sqrt{1+r_{12}^2+r_{13}^2+r_{23}^2}$.

The procedure can be generalized to the states (\ref{GPS-general}) by taking care of two steps:
first consider the base state ${\ket{\mathcal{C}}}={\ket{n_1,n_2,...,n_L}}$ then the base state in the domain wall of the $\sigma^x$ is going to be:

\begin{eqnarray}{\label{base state of domain wall}}
{\ket{\tilde{\mathcal{C}}}}=\begin{cases}
{\ket{n_1,n_2,...,n_L}},\hspace{1.8cm} \sum_{j=1}^Ln_j\ \text{is even},\\
{\ket{n_1,n_2,...,|n_L-1|}},\hspace{1cm} \sum_{j=1}^Ln_j\ \text{is odd}.
\end{cases}
\end{eqnarray}
The two sign sequences in the $\sigma^x$  basis that have the same domain wall configurations do not have the same sign in general. The signs can be found using the following procedure: we first associate to the sign $s_j=\pm$ the number $\bar{s}_j=\pm1$ then the sign for the sequence ${\ket{s_1,s_2,...,s_L}}$ can be determined from the following formula:

\begin{equation}{\label{sign of sequence}}
sgn (\mathcal{S})=\prod_{j=1}^L(-1)^{(n_j-1)(\frac{\bar{s}_j-1}{2})}.
\end{equation}
An example of $L=3$ is

\begin{eqnarray}{\label{GPS-example-sigmax}}
    \ket{\bold{R},101}&=&\frac{1}{\sqrt{2}\mathcal{Z}}
    \Big{(}d\ket{+++}+c\ket{++-}-b\ket{+-+}\nonumber\\
    &-&a\ket{+--}+a\ket{-++}+b\ket{-+-}\nonumber\\&-&c\ket{--+} 
    -d\ket{---}    \Big{)},
\end{eqnarray}
For more examples, see the Appendix \ref{sec:AppExamplesPure}.

\subsection{Gaussian pure states in the $\sigma^x$ basis: periodic boundary condition}\label{sec:GPSsigmaxPBC}

The procedure outlined in the previous section is fairly general, and it works for any system independent of dimension and boundary conditions. However, when we have a one-dimensional periodic boundary condition one can significantly simplify the equations. In this case, the anti-symmetric matrix $R$ is also anti-circulant. In other words, we have

\begin{equation}
    \begin{split}
\bold{R}= \left(
\begin{array}{cccccccc}
 0 & r_{12} & r_{13} & \hdots & r_{1,\frac{L}{2}+1} & r_{1,\frac{L}{2}} & \hdots & r_{12} \\
 -r_{12} & 0 & r_{12} & \hdots & r_{1,\frac{L}{2}+1} & r_{1,\frac{L}{2}} & \hdots & r_{13} \\
 -r_{13} & -r_{12} & 0 & r_{12} & r_{13} & r_{14} & \hdots & r_{14} \\
\vdots &\vdots &\vdots &\vdots &\vdots &\vdots &\ddots &\vdots \\
  -r_{13} & -r_{14} & \hdots & -r_{1,\frac{L}{2}+1} & -r_{1,\frac{L}{2}} &\hdots & 0 & r_{12}\\
 -r_{12} & -r_{13} & \hdots & -r_{1,\frac{L}{2}+1} & -r_{1,\frac{L}{2}} &\hdots & -r_{12} & 0 \\
\end{array}
\right),\hspace{2cm} L\ is\ even
\end{split}
\end{equation}

and 

\begin{equation}\label{eq: M matrix}
\begin{split}
\bold{R}= \left(
\begin{array}{cccccccccc}
 0 & r_{12} & r_{13} & \hdots & r_{1,[\frac{L}{2}]+1} & r_{1,[\frac{L}{2}]+1} & r_{1,[\frac{L}{2}]} & \hdots & r_{12} \\
 -r_{12} & 0 & r_{12} & \hdots & r_{1,[\frac{L}{2}]+1} & r_{1,[\frac{L}{2}]+1} & r_{1,[\frac{L}{2}]} & \hdots & r_{13} \\
\vdots &\vdots &\vdots &\vdots &\vdots &\vdots &\vdots &\ddots &\vdots\\
  -r_{13} & -r_{14} & \hdots& -r_{1,[\frac{L}{2}]+1} & -r_{1,[\frac{L}{2}]+1} & -r_{1,[\frac{L}{2}]} &\hdots & 0 & r_{12}\\
 -r_{12} & -r_{13} & \hdots& -r_{1,[\frac{L}{2}]+1} & -r_{1,[\frac{L}{2}]+1} & -r_{1,[\frac{L}{2}]} &\hdots & -r_{12} & 0 \\
\end{array}
\right), L\ is\ odd
\end{split}
\end{equation}

To make a more  general statement, consider the following skew-circulant matrix
\begin{equation}
\bold{C} = \left(
\begin{array}{ccccc}
c_0 & c_1 & c_2 & \dots & c_{L-1} \\
-c_{L-1} & c_0 & c_1 & \dots & c_{L-2} \\
-c_{L-2} & -c_{L-1} & c_0 & \dots & c_{L-3} \\
\vdots & \vdots & \vdots & \ddots & \vdots \\
-c_1 & -c_2 & -c_3 & \dots & c_0 \\
\end{array}
\right).
\end{equation}
The above matrix can be diagonalized using the following matrix
\begin{equation}
    \bold{U} =
\begin{cases}
  (y^{(0)}, y^{(L-1)},y^{(1)}, y^{(L-2)},\dots, y^{(\frac{L}{2}-1)}, y^{(\frac{L}{2})}) & \text{L even}, \\
  (y^{(0)}, y^{(L-1)},y^{(1)}, y^{(L-2)},\dots, y^{(\frac{L-1}{2})})& \text{L odd}, \\
\end{cases}
\end{equation}
where $\bold{y}^{(m)}$ reads as
\begin{equation}
\bold{y}^{(m)} = \frac{1}{\sqrt{L}}
    \begin{pmatrix}
    1 \\
    e^{-\frac{2\pi i}{L}(m+\frac{1}{2})} \\
    \vdots \\
    e^{-\frac{2\pi i}{L}(m+\frac{1}{2})(L-1)}
\end{pmatrix}
, \ m = 0, 1, 2, \hdots, L-1.
\end{equation}
Then we have $\bold{C} = \bold{U}\cdot \Lambda^c \cdot\bold{U}^{\dagger}$ where 
\begin{equation}
\Lambda^c = 
    \begin{pmatrix}
        \Lambda_0 & 0 & 0&0&\dots & 0 \\
        0 & \Lambda_{L-1} & 0&0&\dots & 0 \\
        0 & 0 &\Lambda_{1}& 0&\dots & 0 \\
        0 & 0 & 0& \Lambda_{L-2}& \dots & 0 \\
        \vdots & \vdots & \vdots & \vdots & \ddots & \vdots \\
        0 & 0 &0 & 0 & \dots & \Lambda_{\frac{L}{2}}(\Lambda_{\frac{L-1}{2}})
    \end{pmatrix},
\end{equation}
and
\begin{equation}
    \Lambda_m = \sum_{k=0}^{L-1} c_k e^{-\frac{2\pi i k}{L}(m+\frac{1}{2})}.
\end{equation}
Since both $\bold{R}$ and $\bold{P}$ are skew-circulant matrices, we now write
\begin{equation}
    \Tilde{\bold{R}} = \bold{U} \cdot\Lambda^{(\Tilde{\bold{R}})} \cdot\bold{U}^{\dagger},
\end{equation}
\begin{equation}
    \Lambda^{(\Tilde{R})} = \frac{\Lambda^{(R)}\Lambda^{(P)} + \Lambda^{(R)} - \Lambda^{(P)} + \bold{I}}{\Lambda^{(R)}\Lambda^{(P)} - \Lambda^{(R)} -\Lambda^{(P)}- \bold{I}}.
\end{equation}
The next step is finding the pfaffinho's of $\Tilde{\bold{R}}$, i.e. $\bold{pf}\ \Tilde{\bold{R}}$. We first write the $\Tilde{\bold{R}}$ matrix in standard spectral
form of skew-symmetric matrices as follows:

\begin{equation}
    \Tilde{\bold{R}} = \bold{V} \cdot\Sigma \cdot\bold{V}^T, \hspace{1cm}
    \bold{V} = e^\frac{-i\pi}{4} \bold{U} \cdot\bold{K}^{\dagger},
\end{equation}
and 
\begin{equation}
    \bold{K} =
\begin{cases}
  \begin{pmatrix}
        \bold{J} & 0 & 0&\dots & 0 \\
        0 & \bold{J} & 0&\dots & 0 \\
        \vdots & \vdots & \ddots & \vdots & \vdots \\
        0 & 0 &0 & \dots & \bold{J}
    \end{pmatrix} & \text{L even}, \\
    
  \begin{pmatrix}
        \bold{J} & 0 & 0&\dots & 0 \\
        0 & \bold{J} & 0&\dots & 0 \\
        \vdots & \vdots  & \ddots & \vdots & \vdots \\
        0 & 0 &0 & \dots & \bold{1}
    \end{pmatrix} & \text{L odd}. \\
\end{cases}
\end{equation}
Here $\bold{J} = \frac{1}{\sqrt{2}}\begin{pmatrix}
        1 & i  \\
        i & 1  \\
    \end{pmatrix} $ and 
\begin{equation}
    \Sigma =
  \begin{pmatrix}
        0 & i \Tilde{\Lambda}_0 & 0&0&\dots & 0 \\
        -i \Tilde{\Lambda}_0 & 0 & 0&0&\dots & 0 \\
        0 & 0& 0&  i \Tilde{\Lambda}_1 & \dots&0& \\
        0 & 0 & -i \Tilde{\Lambda}_1&0&\dots & 0 \\
        \vdots & \vdots & \vdots & \ddots & \ddots & \vdots \\
    \end{pmatrix}. \\
\end{equation}
Then we use the following theorem
\begin{equation}\label{pf PBC}
    \bold{pf}\ \Tilde{R}_I = \sum_J det V_{IJ}\  \bold{pf}\ \Sigma_{JJ},
\end{equation}
which is valid for any anti-symmetric matrices and can be proved using Berezin integrals over Grassmann numbers. Note that here the matrix $\bold{V}$ is independent of the form of the matrix $\Tilde{\bold{R}}$, which significantly simplifies the calculations of the pfaffinhos of the matrix $\tilde{\bold{R}}$.\\
It is worth mentioning that the amplitude of the configuration $++...+$ can be easily found to be 
\begin{equation}\label{all up}
    \frac{1}{\sqrt{2}}\prod_{j=0}^{L-1}\frac{1}{(1+|\tilde{\Lambda}_j|^2)^{\frac{1}{4}}}.
\end{equation}
Explicit equations like the above can be found for the amplitudes of the configurations that have some periodic structures making the equation (\ref{all up}) amenable to analytical calculations.

%The above procedure gives the amplitudes in the $\sigma^x$ basis to calculate the probabilities we can define the following block matrices 
%\begin{equation}
%     \bold{\mathbb{R}} :=   
%     \begin{pmatrix}
%        \Tilde{\bold{R}} & 0 \\
%        0 & \Tilde{\bold{R}}^* \\
%    \end{pmatrix} \\ ,\hspace{1cm}
%    \bold{\mathbb{V}} := \begin{pmatrix}
%        \bold{V} & 0 \\
%        0 & \bold{V}^* \\
%    \end{pmatrix} \\,\hspace{1cm}
%    \Sigma := 
%    \begin{pmatrix}
%        \Sigma & 0 \\
%        0 & \Sigma \\
%    \end{pmatrix} \\,
%\end{equation}
%where $\bold{\mathbb{R}} = \mathbb{V} \Sigma \mathbb{V}^T$.
%Then we can use again the Equation \ref{pf} with $\mathbb{I}=\{I, I+L\}$, $\mathbb{J}=\{J, J+L\}$ Then we have 
%\begin{equation}
%    \bold{pf}\ \Tilde{\mathbb{R}}_\mathbb{I} =  \lvert \bold{pf}\ \Tilde{\bold{R}} \rvert^2 = \sum_{\mathbb{J}} det \bold{V}_{\mathbb{I}\mathbb{J}}(\bold{pf}\ \Sigma_{\mathbb{J}})^2 = \sum_J det \bold{V}_{IJ}\ det \bold{V}_{IJ}^*\ det\Sigma_J = \sum_{J} \lvert det \bold{V}_{IJ}\rvert^2 det \Sigma_J.
%\end{equation}
%Which is the desired probability. In general, if $\Sigma$ was not real the above equation would have the following form
%\begin{equation}
%    p_I = \sum_{J} \lvert det \bold{V}_{IJ} \rvert^2 \lvert \bold{pf}\ \Sigma_{J} \rvert^2
%\end{equation}

\section{Gaussian pure states in the $(\phi,\frac{\pi}{2},0)$ basis}\label{sec:phi basis}

The procedure outlined in the previous section can be extended also to any basis in the $xy$ plane. For brevity of the notation we use $\tilde{\bold{R}}^{\phi}\equiv\tilde{\bold{R}}^{(\phi,\frac{\pi}{2},0)}$.
Here we need to define the following operator
\begin{equation}\label{eq: O xy}
   \mathcal{O}_{j,j+1}^{\phi}=(\cos{\phi}\sigma_j^x+\sin{\phi}\sigma_{j}^y)(\cos{\phi}\sigma_{j+1}^x+\sin{\phi}\sigma_{j+1}^y).
\end{equation}
Then much like in the previous section, we expect

\begin{eqnarray}{\label{KW1 phi}}
\bra{0,\bold{R}} \mathcal{O}_{j,j+1}^{\phi}\ket{\bold{R},0}&=&-\bra{0,\tilde{\bold{R}}^{\phi}} \sigma_j^z\ket{\tilde{\bold{R}}^{\phi},0},
\end{eqnarray}
which translates into

\begin{eqnarray}{\label{KW1 phi fermion I}}
\cos^2{\phi}G_{j,j+1}(R)+i\sin{\phi}\cos{\phi}\bar{K}_{j,j+1}(R)-i\sin{\phi}\cos{\phi}K_{j,j+1}(R)+\sin^2{\phi}G_{j+1,j}(R)=1-2C_{j,j}(\tilde{R}^{\phi}),\\
{\label{KW1 phi fermion II}}
\cos^2{\phi}G_{L,1}(R)+i\sin{\phi}\cos{\phi}\bar{K}_{L,1}(R)-i\sin{\phi}\cos{\phi}K_{L,1}(R)+\sin^2{\phi}G_{1,L}(R)=2C_{L,L}(\tilde{R}^{\phi})-1,\hspace{1cm}
\end{eqnarray}

in the fermionic space. Here we have defined the matrices

\begin{eqnarray}{\label{G K bar K}}
G_{jk}(R)&=&\bra{0,\bold{R}} (c_j^{\dagger}-c_j)(c_k^{\dagger}+c_k)\ket{\bold{R},0},\\
K_{jk}(R)&=&\bra{0,\bold{R}} (c_j^{\dagger}+c_j)(c_k^{\dagger}+c_k)\ket{\bold{R},0},\\
\bar{K}_{jk}(R)&=&-\bra{0,\bold{R}} (c_j^{\dagger}-c_j)(c_k^{\dagger}-c_k)\ket{\bold{R},0},
\end{eqnarray}
and the matrix $\boldsymbol{C}$ is defined as the equation (\ref{C defenitions}). The above matrices can be calculated explicitly as:

\begin{eqnarray}{\label{ G and K and bar K}}
\bold{G}&=&\bold{I}+\bold{Q}^T\cdot(\bold{R}-\bold{I})+\bold{R}^*\cdot\bold{Q}^T-\bold{Q},\\
\bold{K}&=&(\frac{\bold{I}}{2}+\bold{R}^*)\cdot\bold{Q}^T+\bold{Q}^T\cdot(\frac{\bold{I}}{2}-\bold{R})-\bold{Q}+\bold{I},\\
\bar{\bold{K}}&=&(\frac{\bold{I}}{2}-\bold{R}^*)\cdot\bold{Q}^T+\bold{Q}^T\cdot(\frac{\bold{I}}{2}+\bold{R})-\bold{Q}+\bold{I}.
\end{eqnarray}
A solution to the equations (\ref{KW1 phi fermion I}) and (\ref{KW1 phi fermion II}) can be found as follows:
First, we make the matrix

\begin{equation}\label{eq: R phi}
   \bold{R}^{\phi}=e^{2i\phi}\bold{R}.
\end{equation}
and then the matrix
\begin{eqnarray}{\label{H phi}}
\bold{H}^{\phi} = (\bold{R}^{\phi}-\bold{I})\cdot(\bold{R}^{\phi}+\bold{I})^{-1}.
\end{eqnarray}
Finally, we have
\begin{equation}{\label{R tilda phi}}
    \Tilde{\bold{R}}^{\phi} = (\bold{I}+\bold{H}^{\phi}\cdot\bold{P})\cdot(\bold{H}^{\phi}\cdot\bold{P}-\bold{I})^{-1}.
\end{equation}
The above matrix can now be used to directly construct the state in the $(\phi,\frac{\pi}{2},0)$ basis as we outlined in the previous section. 

To summarize: we first construct the state 

\begin{equation}{\label{Domainwall0}}
    \ket{\tilde{\bold{R}}^{\phi},\tilde{0}} = \frac{1}{\mathcal{N}_{\tilde{R}^{\phi}}}e^{{\frac{1}{2}\sum_{i,j}^Lc^{\dagger}_{i}(\tilde{R}^{\phi})_{ij}c^{\dagger}_{j}}}\ket{\tilde{0}}.
\end{equation}
Then we assume it describes the $\ket{\bold{R},0} $  in the domain wall basis of the $(\frac{\pi}{2},\phi,0)$ basis. Consequently, the state on the desired basis will be

\begin{equation}{\label{GPS0-xy basis}}
    \ket{\bold{R},0} = \frac{1}{\sqrt{2}\mathcal{N}_{\tilde{R}^{\phi}}}\sum_{\mathcal{S}}sgn(\mathcal{S})\bold{pf}\ (\tilde{R}^{\phi})_{\mathcal{S}}{\ket{\mathcal{S}}}_{\phi},
\end{equation}
where ${\ket{\mathcal{S}}}_{\phi}$ as before is a sequence of $+$ and $-$ in the $\phi$ basis and the $(\tilde{R}^{\phi})_{\mathcal{S}}$ is a submatrix of the matrix $\tilde{\bold{R}}^{\phi}$ in which we first find the domain wall configuration of $\mathcal{S}$ and then we keep the rows and columns corresponding to the sites that there is a domain wall. The $sgn(\mathcal{S})$ is simply +1(-1) for even(odd) number of $-$ in the sequence.

An example of $L=3$ is

\begin{eqnarray}{\label{GPS-example-sigmax}}
    \ket{\bold{R},000}&=&\frac{1}{\sqrt{2}\mathcal{Z}^{\phi}}
    \Big{(}a^{\phi}_{(000)}\ket{+++}-b^{\phi}_{(000)}\ket{++-}\\
    &-&c^{\phi}_{(000)}\ket{+-+}+d^{\phi}_{(000)}\ket{+--}-d^{\phi}_{(000)}\ket{-++}\nonumber\\
    &+&c^{\phi}_{(000)}\ket{-+-}+b^{\phi}_{(000)}\ket{--+} -a^{\phi}_{(000)}\ket{---}    \Big{)}\nonumber,
\end{eqnarray}
where 
\begin{eqnarray}{\label{abcd}}
a^{\phi}_{(000)}&=&1+e^{2i\phi}
(r_{12}+r_{13}+r_{23}),\nonumber\\
b^{\phi}_{(000)}&=&1+e^{2i\phi}
(r_{12}-r_{13}-r_{23}),\nonumber\\
c^{\phi}_{(000)}&=&1+e^{2i\phi}
(-r_{12}+r_{13}-r_{23}),\nonumber\\
d^{\phi}_{(000)}&=&1+e^{2i\phi}
(-r_{12}-r_{13}+r_{23}),\nonumber\\
\mathcal{Z}^{\phi}&=&2\sqrt{1+r_{12}^2+r_{13}^2+r_{23}^2}.\nonumber
\end{eqnarray}

The procedure can be generalized to the states (\ref{GPS-general}) by the method outlined for the $x$ basis with one extra twist:
first consider the base state ${\ket{\mathcal{C}}}={\ket{n_1,n_2,...,n_L}}$, then the base state in the domain wall of the $\phi$ is going to be again:

\begin{eqnarray}{\label{base state of domain wall}}
{\ket{\tilde{\mathcal{C}}}}=\begin{cases}
{\ket{n_1,n_2,...,n_L}},\hspace{1.8cm} \sum_{j=1}^Ln_j\ \text{is even},\\
{\ket{n_1,n_2,...,|n_L-1|}},\hspace{1cm} \sum_{j=1}^Ln_j\ \text{is odd}.
\end{cases}
\end{eqnarray}
As before, the two sign sequences in the $\phi$  basis that have the same domain wall configurations do not have the same sign in general. The signs can be found using a similar procedure as before: we first associate to the sign $s_j=\pm$ the number $\bar{s}_j=\pm1$ then the sign for the sequence ${\ket{s_1,s_2,...,s_L}}$ can be determined from the following formula:

\begin{equation}{\label{sign of sequence}}
sgn (\mathcal{S})=\prod_{j=1}^L(-1)^{(n_j-1)(\frac{\bar{s}_j-1}{2})}.
\end{equation}
The twist comes in the determination of the matrix $\bold{R}_{\phi}$. The elements of this matrix are

\begin{equation}\label{eq: R phi}
   r_{i,j}^{\phi}=r_{i,j}e^{2i\phi(1-n_i-n_j)}.
\end{equation}
After making the matrix $\bold{R}^{\phi}$ one can use the equation (\ref{R tilda phi}) to write the state in the desired basis.

An example of $L=3$ is

\begin{eqnarray}{\label{GPS-example-sigmax}}
    \ket{\bold{R},101}&=&\frac{1}{\sqrt{2}\mathcal{Z}^{\phi}}
    \Big{(}d^{\phi}_{(101)}\ket{+++}+c^{\phi}_{(101)}\ket{++-}\nonumber\\
    &-&b^{\phi}_{(101)}\ket{+-+}-a^{\phi}_{(101)}\ket{+--}\nonumber\\
    &+&a^{\phi}_{(101)}\ket{-++}+b^{\phi}_{(101)}\ket{-+-}\nonumber\\
    &-&c^{\phi}_{(101)}\ket{--+} -d^{\phi}_{(101)}\ket{---}    \Big{)},
\end{eqnarray}
where 
\begin{eqnarray}{\label{abcd}}
a^{\phi}_{(101)}&=&1+r_{12}+e^{-2i\phi}r_{13}+r_{23},\nonumber\\
b^{\phi}_{(101)}&=&1+r_{12}-e^{-2i\phi}r_{13}-r_{23},\nonumber\\
c^{\phi}_{(101)}&=&1-r_{12}+e^{-2i\phi}r_{13}-r_{23},\nonumber\\
d^{\phi}_{(101)}&=&1-r_{12}-e^{-2i\phi}r_{13}+r_{23}.\nonumber
\end{eqnarray}

\section{Gaussian pure states in the $(\phi,\frac{\pi}{2},\alpha)$ basis}\label{sec:phi-alpha basis}

Unlike the previous sections, the Gaussian states in the  $(\phi,\frac{\pi}{2},\alpha)$ basis when $\alpha\neq0$ does not have a domain wall kind of interpretation. We conjecture that it can be obtained first by following the steps that lead the state in the $(\phi,\frac{\pi}{2},0)$ basis and then just putting $e^{-i\alpha}$ for each $-$ in the corresponding configuration. For example, we have
\begin{eqnarray}{\label{GPS-example-sigmax}}
    \ket{\bold{R},101}&=&\frac{1}{\sqrt{2}\mathcal{Z}^{\phi}}
    \Big{(}d^{\phi}_{(101)}\ket{+++}+e^{i\alpha}c^{\phi}_{(101)}\ket{++-}\nonumber\\
    &-&e^{-i\alpha}b^{\phi}_{(101)}\ket{+-+}-e^{-2i\alpha}a^{\phi}_{(101)}\ket{+--}\nonumber\\
    &+&e^{-i\alpha}a^{\phi}_{(101)}\ket{-++}+e^{-2i\alpha}b^{\phi}_{(101)}\ket{-+-}\nonumber\\
    &-&e^{-2i\alpha}c^{\phi}_{(101)}\ket{--+} \nonumber\\
    &-&e^{-3i\alpha}d^{\phi}_{(101)}\ket{---}\Big{)}.
\end{eqnarray}
Note that the probability of the amplitudes are independent of the phase $\alpha$.

\section{Configuration probabilities}\label{sec:probabilities}
In this segment, we will present a set of formulas designed to efficiently calculate the probability of each configuration within the basis outlined in the preceding section. We will focus on the specific case mentioned as (\ref{GPS0}), but it's worth noting that analogous formulas can be readily developed for more general scenarios as well. Subsequently, we will apply these formulas to analyze the probabilities of formation in the transverse field Ising chain at its critical point, revealing intriguing universal behaviors.

We start with the computational basis, i.e. $\sigma^z$ basis. It is easy to see that for the configuration $\mathcal{I}$, we will have
\begin{equation}{\label{Prozgeneric}}
    P_{\mathcal{I}} = \frac{|\bold{pf}R_{\mathcal{I}}|^2}{\mathcal{N}_R^2}.
\end{equation}
When the matrix $\bold{R}$ is real it simplifies to \cite{MortezaRajab2020}
\begin{equation}{\label{Prozreal}}
    P_{\mathcal{I}} = \det[\frac{\bold{I}-\bold{I}_{\mathcal{I}}\cdot\bold{G}}{2}],
\end{equation}
where the matrix $\bold{I}_{\mathcal{I}}$ is diagonal, composed of $\pm1$, with its composition clearly influenced by the specific columns and rows that are eliminated. We assign a diagonal element of $-1$ in cases where a fermion is present, and $1$ in instances where there is an absence of a fermion at the relevant site.

The next step is writing the probabilities for the bases in the $xy$ plane. Using the amplitude formulas of the previous sections and some simplifications for the probability of the configuration $\mathcal{S}$, we have
\begin{equation}{\label{Prophigeneric}}
    P^{\phi}_{\mathcal{S}} =\frac{1}{2} \frac{|\det \tilde{R}^{\phi}_{\mathcal{S}}|}{\mathcal{N}_{\tilde{R}^{\phi}}^2},
\end{equation}
where 

\begin{equation}{\label{Rtildaphi}}
 \tilde{R}^{\phi}=\Big{(}\bold{H}-i\cot{\phi}\bold{H}\cdot\bold{P}+\bold{P}-i\cot{\phi}\bold{I}\Big{)}\cdot\Big{(}-\bold{H}-i\cot{\phi}\bold{H}\cdot\bold{P}+\bold{P}+i\cot{\phi}\bold{I}\Big{)}^{-1}   
\end{equation}

The equation \ref{Prophigeneric} can be also written as
\begin{equation}{\label{Prophigeneric2}}
    P^{\phi}_{\mathcal{S}} =\frac{1}{2\mathcal{N}_{\tilde{R}^{\phi}}^2} |\det[\frac{(\bold{I}-\bold{I}_{\mathcal{S}})\cdot\tilde{\bold{R}}^{\phi}+\bold{I}+\bold{I}_{\mathcal{S}}}{2}]|,
\end{equation}
where the diagonal matrix $\bold{I}_{\mathcal{S}}$ is defined for the configuration $\mathcal{S}$ by first finding the domain wall configuration, and then we put $-1(+1)$ in the diagonal of the matrix $\bold{I}_{\mathcal{S}}$ when there is (no) domain wall at the site.

The above formula can be further simplified when the $\bold{R}$ matrix is real and we are interested in the $\sigma^x$ or $\sigma^y$ bases as follows:
\begin{eqnarray}{\label{Proxandy}}
    P^x_{\mathcal{S}} =&\frac{1}{2} \det[\frac{\bold{I}-\bold{I}_{\mathcal{S}}\cdot\bold{G}\cdot\bold{P}}{2}],\\
    P^y_{\mathcal{S}} =&\frac{1}{2} \det[\frac{\bold{P}-\bold{I}_{\mathcal{S}}\cdot\bold{G}}{2}].
\end{eqnarray}
Interestingly, the aforementioned formulas eliminate the need to invert any matrix, thereby facilitating efficient computation of probabilities for large-scale systems.

\subsection{Formation probabilities in the critical transverse field Ising chain}

In this section, we explore the formation probabilities for the ground state of the transverse field Ising (TFI) chain at its critical point. We consider both periodic and open boundary conditions.

For systems like the TFI chain under critical conditions, the logarithm of formation probabilities can be interpreted as the free energy in a Conformal Field Theory (CFT) setting. Specifically, when periodic boundary conditions are used, the system resembles a CFT on a cylinder, with boundaries at both ends. Under these conditions, an additional term known as boundary entropy emerges, contributing to the bulk free energy. Conversely, when open boundary conditions are applied, the system transforms into what can be visualized as a strip. This configuration imposes a natural boundary condition that directly arises from the Hamiltonian’s properties. In such cases, the boundary conditions determined by the formation probabilities might align with or deviate from the system’s natural boundary conditions. This deviation sometimes requires the use of conformal boundary changing operators. For further details and discussion, see references~\cite{AL1991,Stephan2009,Rajabpour2020}.

Based on the above arguments coming from CFT, one would anticipate the formation probabilities to exhibit the following behavior( for those configurations that converge toward conformal boundary conditions) for the Periodic Boundary Condition (PBC) \cite{AL1991,Stephan2009,Rajabpour2020}:
\begin{eqnarray}{\label{CFT-expectationPBC}}
   -\ln P_{\mathcal{S}} =\Gamma_{\mathcal{S}}L-2s_{\mathcal{S}}+\mathcal{O}(\frac{1}{L}),
\end{eqnarray}
where $s_{\mathcal{S}}$ is a universal quantity and is dubbed as the Affleck-Ludwig boundary entropy. It should be noted that for all systems belonging to the Ising universality class, such as the XY chain at the critical Ising line ($h=1$), we anticipate that this quantity will be universal and take one of the three values that we will list later. For detailed study in the $\sigma^z$ basis, see \cite{Rajabpour2020}. In addition, in this study, we show that the quantity is also universal with respect to the change of basis.

In the case of Open Boundary Condition (OBC), one expects a totally different behavior \cite{Peschel1988,Stephan2013}
\begin{eqnarray}{\label{CFT-expectationOBC}}
   -\ln P_{\mathcal{S}} =\Gamma_{\mathcal{S}}L+a_{\mathcal{S}}\ln L+\mathcal{O}(1),
\end{eqnarray}
where this time the coefficient $a_{\mathcal{S}}$ is the universal quantity which depends on the central charge $c$ and the conformal weight of the boundary changing operator  $h_{bcc}$ as follows:
\begin{eqnarray}{\label{coefficient-a}}
  a_{\mathcal{S}}=8h_{bcc}-\frac{c}{4}.
\end{eqnarray}
When the configuration flows to the boundary condition compatible with the natural boundary condition of the system, then  $h_{bcc}=0$; otherwise, depending on the induced boundary condition, one needs to add the corresponding conformal weight in the above formula.

We will analyze the aforementioned formulas for the ground state of the TFI chain across all bases that allow us to study large sizes and various configurations. In the majority of scenarios, we will be able to numerically determine the universal quantities without any ambiguity.

The Hamiltonian of the TFI model is defined as follows:
\begin{equation}{\label{Ising Hamiltonian}}
   H=-\frac{1}{2}\sum_{j=1}^{L'}\sigma_j^x\sigma_{j+1}^x-\frac{h}{2}\sum_{j=1}^L\sigma_j^z,
\end{equation}
where $L'=L$ for PBC with $\sigma_{L+1}^{x}=\sigma_1^x$ and $L'=L-1$ for OBC. 

When $L$ is even the ground state of the critical TFI chain, i.e. $h=1$, has the form (\ref{GPS0}) with the following $\bold{R}$ matrix:
\begin{eqnarray}{\label{Rmatrix}}
 \bold{R}=(\bold{I}+\bold{G})\cdot(\bold{I}-\bold{G})^{-1},
\end{eqnarray}
where 
\begin{equation}{\label{boundary entropy}}
 G_{nm}^{pbc}=\frac{(-1)^{n-m}}{L\sin\frac{\pi(n-m+1/2)}{L}},
\end{equation}
\begin{equation}{\label{boundary entropy}}
 G_{nm}^{obc}=\frac{(-1)^{n-m}}{2L+1}\Big{(}\frac{1}{\sin\frac{\pi(n-m+1/2)}{2L+1}}+\frac{1}{\sin\frac{\pi(n+m-1/2)}{2L+1}}\Big{)}.
\end{equation}

A configuration in the TFI chain may converge to one of three potential conformal boundary conditions: free, fixed, and mixed, each associated with the following boundary entropies\cite{Konechny2017,Rajabpour2020}:
\begin{eqnarray}{\label{boundary entropyIsing}}
    s_{\mathcal{S}}=\begin{cases}
-\frac{\ln2}{2},&\hspace{1.8cm} \text{fixed},\\
0,&\hspace{1.8cm} \text{free},\\
\frac{\ln2}{2},&\hspace{1.8cm} \text{mixed},
\end{cases}
\end{eqnarray}
and the log coefficients in the OBC:
\begin{eqnarray}{\label{boundary entropy}}
    a_{\mathcal{S}}=\begin{cases}
\frac{3}{8},&\hspace{1.8cm} \text{fixed},\\
-\frac{1}{8},&\hspace{1.8cm} \text{free},\\
\frac{3}{8},&\hspace{1.8cm} \text{mixed}.
\end{cases}
\end{eqnarray}
In deriving above we used $c=\frac{1}{2}$ and $h_{bcc}=\frac{1}{16}$ for the fixed and mixed boundary conditions and $h_{bcc}=0$  for free boundary conditions in the equation (\ref{coefficient-a}). Note that the natural boundary condition of the Hamiltonian with OBC is compatible with the free BC.

The configurations under investigation are those possessing a crystalline structure based on a foundational configuration. For instance, take a base configuration of size $p$, within which $u$ elements are oriented upwards. The complete configuration then consists of this base pattern repeated $n$ times, ensuring the total system size, $L$, is an even number. Configurations of this nature have been explored in the $\sigma^z$ basis using both numerical and analytical methods, as documented in the references (\cite{NR2016,Rajabpour2020}). In the case of PBC, the analytical calculations in (\cite{Rajabpour2020}) reveal the following pattern for the boundary entropy:
\begin{eqnarray}{\label{boundary entropy crystal}}
    s_{\mathcal{S}}=\begin{cases}
0,\hspace{2cm} \text{$p-u$ even},\\
\frac{\ln2}{2},\hspace{1.8cm} \text{$p-u$ odd}.
\end{cases}
\end{eqnarray}
For OBC, after considering various configurations, we found that
\begin{eqnarray}{\label{log coefficient entropy crystal}}
    a_{\mathcal{S}}=\begin{cases}
-\frac{1}{8},\hspace{1.6cm} \text{$p-u$ even},\\
\frac{3}{8},\hspace{1.8cm} \text{$p-u$ odd}.
\end{cases}
\end{eqnarray}
The above conclusion is consistent with what we expect from boundary CFT. In our analysis, we considered systems of sizes as large as $L\approx 1000$, focusing our data fitting primarily on sizes within the range $(400 \leq L \leq 1000)$. The same system sizes were employed across all other bases and configurations. It's worth mentioning that the choice of $L\approx 1000$ was not due to computational limitations but rather because the fit quality is already excellent at this size. Furthermore, since the calculations involve computing a series of determinants, it is feasible to extend our analysis to system sizes of several thousand using Mathematica with relative ease. The numbers presented were obtained by fitting the logarithm of formation probabilities to equations (\ref{CFT-expectationPBC}) and (\ref{CFT-expectationOBC}) using Mathematica. The outcomes of the fitting process align within less than one percent of the values that have been reported.

In the next step we studied the configurations in the $(\phi,\frac{\pi}{2},0)$ basis when $\phi\ne\frac{\pi}{2}$. Remarkably we found the following universal behavior for the boundary entropy:
\begin{eqnarray}{\label{boundary entropy crystal phi basis}}
    s_{\mathcal{S}}=\begin{cases}
0,\hspace{2.2cm} \text{$p=2u$ },\\
-\frac{\ln2}{2},\hspace{1.8cm} \text{$p\neq2u$},
\end{cases}
\end{eqnarray}
independent of the value of $\phi$. For OBC, the coefficient of the logarithm shows the following universal behavior:
\begin{eqnarray}{\label{log coefficient entropy crystal phi basis}}
    a_{\mathcal{S}}=\begin{cases}
-\frac{1}{8},\hspace{1.6cm} \text{$p=2u$},\\
\frac{3}{8},\hspace{1.8cm} \text{$p\neq2u$}.
\end{cases}
\end{eqnarray}
Again independent of the angle $\phi$.

We also studied the configurations in the $\sigma^y$ basis. In this case, as far as $p=2u$ the results are the same as the general $\phi$ basis. However, when $p\neq2u$ there are various irregularities with respect to the size of the system $L$ that we do not understand. When $|\mathcal{S}\rangle=|+++..\rangle$  our numerical results indicate the following boundary entropies:
\begin{eqnarray}{\label{boundary entropy crystal y basis all up}}
    s_{\mathcal{S}}=\begin{cases}
+\frac{\ln2}{2},\hspace{2.2cm} \text{$L=6k$ },\\
-\frac{\ln2}{2},\hspace{1.8cm} \text{$L=6k+2,6k+4$},
\end{cases}
\end{eqnarray}
where $k$ is an integer number. For OBC we found $a=\frac{3}{8}$ when $L=6k+6$. In the case of the base configuration $++-$, we were not able to find a consistent result that pinpointed the relevant system sizes. However, surprisingly in the case of the base configuration $+++-$, we found 
\begin{eqnarray}{\label{boundary entropy crystal y basis uuud}}
    s_{\mathcal{S}}=\begin{cases}
+\frac{\ln2}{2},\hspace{2.2cm} \text{$L=72k+72$ },\\
-\frac{\ln2}{2},\hspace{1.8cm} \text{$L=72k+24,72k+48$}. 
\end{cases}
\end{eqnarray}
The coefficient of the logarithm in the case of OBC for the above sizes, i.e. $L=72k+24,72k+48,72k+72$, is $a=\frac{3}{8}$. 

We also tried other configurations with $p\neq2u$ but were not able to find any pattern in these cases.

\section{Conclusions}\label{sec:Conclusion}

In our study, we have developed a precise and explicit formula for determining the amplitudes of arbitrary Gaussian pure states within the $(\phi,\frac{\pi}{2},\alpha)$ basis. This formula expresses the amplitudes through the Pfaffian of a submatrix derived from a clearly defined matrix. The significance of this development is that it allows for the computation of amplitudes for specific configurations in relatively large systems, owing to the polynomial computational complexity of calculating the Pfaffian. Particularly in periodic systems, which result in anticirculant matrices, our method can yield explicit analytical formulas.

However, this paper does not address scenarios where $\theta \neq \frac{\pi}{2},0$. The approach using domain wall configurations seems inadequate, or perhaps there is an aspect we have not yet identified. Extending this methodology to cover these cases would complete the narrative of determining Gaussian pure states on any arbitrary basis.

Using our formulas, we analyzed the formation probabilities for various configurations in the ground state of the critical transverse field Ising chain, considering both periodic and open boundary conditions. This examination across different configurations revealed a coherent understanding of the universal quantities, such as boundary entropy in the periodic case and the logarithm coefficient in the open case, for all bases in the $xy$ plane except the $\sigma^y$ basis. In this particular basis, we noted significant anomalies for crystalline configurations with unequal numbers of up and down spins. Investigating these specific configurations through analytical methods, especially for periodic chains, could provide a comprehensive insight into the phenomena.
In the future, we plan to delve into Gaussian mixed states to further enhance our understanding and expand the practical applications of Gaussian states in quantum systems.
%Looking ahead, the natural progression of this research lies in exploring Gaussian mixed states, a topic we aim to revisit in future work. Such an extension would deepen the understanding and broaden the applicability of Gaussian states in quantum systems.%

\section*{Acknowledgements}
We thank M. Heyl for reading the manuscript and comments.  We thank  M. A. Seifi Mirjafarlou for pointing us to a few typos. MAR thanks CNPq and FAPERJ (grant number E-26/210.062/2023) for partial support.

%\newpage

\appendix

%\begin{appendices}
\addtocontents{toc}{\protect\setcounter{tocdepth}{1}}
\renewcommand{\theequation}{\thesection.\arabic{equation}}

\setcounter{equation}{0}
\section{Generic Gaussian pure states}\label{sec:GaussianStateGeneral}
In this appendix, we will show that more general Gaussian states can be written in the form (\ref{GPS-general}).
Consider the following generic Gaussian pure state:
\begin{equation}\label{eq: F}
\begin{split}
e^{ \frac{1}{2} \begin{pmatrix} \mathbf{c}^{\dagger} & \mathbf{c}\\ \end{pmatrix} \mathbf{M} \begin{pmatrix} \mathbf{c} \\ \mathbf{c}^{\dagger}  \end{pmatrix}}\ket{\mathcal{C}},
\end{split}
\end{equation}
where $\left(\mathbf{c}^{\dagger}, \mathbf{c}\right)=\left(c_1^{\dagger},c_2^{\dagger}, ... , c_L^{\dagger},c_1,c_2, ... , c_L\right)$. Without losing generality, it is a requirement that $\mathbf{J} .\mathbf{M} $ be an antisymmetric matrix, where the $\mathbf{J}$ matrix is defined as follows:
\begin{equation}
\begin{split}
\mathbf{J} = \begin{pmatrix} \mathbf{0} & \mathbf{I} \\ \mathbf{I} & \mathbf{0} \end{pmatrix}.
\end{split}
\end{equation}
In this Appendix, we will show that the above state can be written as the equation (\ref{GPS-general}).

We first do the following canonical transformation:

\begin{equation}\label{eq: transformation}
\begin{split}
e^{ \frac{1}{2} \begin{pmatrix} \mathbf{c}^{\dagger} & \mathbf{c}\\ \end{pmatrix} \boldsymbol{\Pi}^2_{\mathcal{C}} \mathbf{M} \boldsymbol{\Pi}^2_{\mathcal{C}} \begin{pmatrix} \mathbf{c} \\ \mathbf{c}^{\dagger}  \end{pmatrix}}\ket{\mathcal{C}}\to e^{ \frac{1}{2} \begin{pmatrix} \bar{\mathbf{c}}^{\dagger} & \bar{\mathbf{c}}\\ \end{pmatrix} \bar{\mathbf{M}}_{\mathcal{C}} \begin{pmatrix} \bar{\mathbf{c}} \\ \bar{\mathbf{c}}^{\dagger}  \end{pmatrix}}\ket{\bar{0}},
\end{split}
\end{equation}
where the permutation matrix $\boldsymbol{\Pi}_{\mathcal{C}}$ is made of the multiplication of the following canonical permutation matrices

\begin{equation}
    \boldsymbol{\Pi}_{j}=\begin{pmatrix}
        \mathbf{I}_j & \mathbf{O}_j\\
        \mathbf{O}_j & \mathbf{I}_j\\
    \end{pmatrix},
    \quad\text{where}\quad
    \begin{cases}
                (\mathbf{I}_{j})_{nm}=\delta_{n,m}(1-\delta_{n,j}), & \\
                (\mathbf{O}_{j})_{nm}=\delta_{n,m}\delta_{n,j}, & \\
                 \end{cases}
\end{equation}
in which $\boldsymbol{\Pi}_{j}:\;c_j^\dagger\rightleftarrows c_{j}^{}$. Then $\boldsymbol{\Pi}_{\mathcal{C}}$ is chosen such that we do the exchange $c_j^\dagger\rightleftarrows c_{j}$ for all the sites that there is a fermion. In this way we have $\bar{c}_i\ket{\bar{0}}=0, \forall i$. The next step is using Balian-Brezin decomposition such that

\begin{equation}\label{eq: balian-Brezin}
\begin{split}
 &e^{ \frac{1}{2} \begin{pmatrix} \bar{\mathbf{c}}^{\dagger} & \bar{\mathbf{c}}\\ \end{pmatrix} \bar{\mathbf{M}}_{\mathcal{C}} \begin{pmatrix} \bar{\mathbf{c}} \\ \bar{\mathbf{c}}^{\dagger}  \end{pmatrix}}\ket{\bar{0}}=\\ &e^{\frac{1}{2}\bar{\mathbf{c}}^{\dagger} \bar{\mathbf{R}}\bar{\mathbf{c}}^{\dagger}} e^{\bar{\mathbf{c}}^{\dagger} \bar{\mathbf{Y}}\bar{\mathbf{c}}-\frac{1}{2} Tr\bar{\mathbf{Y}}}e^{\frac{1}{2}\bar{\mathbf{c}} \bar{\mathbf{Z}}\bar{\mathbf{c}}}\ket{\bar{0}}\to e^{\frac{1}{2}\bar{\mathbf{c}}^{\dagger} \bar{\mathbf{R}}\bar{\mathbf{c}}^{\dagger}}\ket{\bar{0}},
\end{split}
\end{equation}

where
 \begin{equation}\label{eq: T_22 XYZ}
\begin{split}
& \bar{\mathbf{R}}=\bar{\mathbf{T}}_{12} (\bar{\mathbf{T}}_{22})^{-1}, \ \ \ \ \ \bar{\mathbf{Z}}=(\bar{\mathbf{T}}_{22})^{-1}\bar{\mathbf{T}}_{21},  \ \ \ \ \ e^{\mathbf{-\bar{Y}}}=\bar{\mathbf{T}}_{22}^T.
\end{split}
\end{equation}
and
\begin{equation}
\begin{split}
 \bar{\mathbf{T}} = \begin{pmatrix} \bar{\mathbf{T}}_{11} & \bar{\mathbf{T}}_{12} \\ \bar{\mathbf{T}}_{21} & \bar{\mathbf{T}}_{22} \end{pmatrix}=e^{\bar{\mathbf{M}}}.
\end{split}
\end{equation}

After writing back the state (\ref{eq: balian-Brezin}) with respect to the original creation and annihilation operators, the above state has the form (\ref{GPS-general}).

\setcounter{equation}{0}
\section{Examples of Gaussian pure states}\label{sec:AppExamplesPure}
\setcounter{equation}{0}
In this Appendix, we list some examples of the Gaussian states in different bases for various sizes. We first list all the possibilities for the $L=2$, and then for $L=3$ we just list them for the $\sigma^z$ and $\sigma^x$ bases.

\subsection{Gaussian pure states with size $L = 2$}

In this section, we write the Gaussian states with $L=2$ in different bases. We first define:

\begin{eqnarray}{\label{abcd}}
a^{\phi}_{(00)}&=&1+e^{2i\phi}r_{12},\ 
a^{\phi}_{(01)}\ =\ 1+r_{12},\
a^{\phi}_{(10)}\ =\ 1+r_{12},\
a^{\phi}_{(11)}\ =\ 1+e^{-2i\phi}r_{12},\nonumber\\
b^{\phi}_{(00)}&=& 1-e^{2i\phi}r_{12},\ 
b^{\phi}_{(01)}\ =\ 1-r_{12},\ 
b^{\phi}_{(10)}\ =\ 1-r_{12},\ 
b^{\phi}_{(11)}\ =\ 1-e^{-2i\phi}r_{12},\nonumber\\
\mathcal{N}_R&=& \sqrt{1+r_{12}^2},\nonumber\\
\mathcal{Z}^{\phi}&=&2\sqrt{1+r_{12}^2},\nonumber
\end{eqnarray}

and 
\begin{equation}
    \textbf{R} = \left(
\begin{array}{cc}
 0 & \text{$r_{12}$}  \\
 -\text{$r_{12}$} & 0  \\
\end{array}
\right).
\end{equation}
The corresponding  $\bold{R}^{\phi}$ and $\Tilde{\bold{R}}^{\phi}$ are listed in the Table \ref{tab:matrix_comparison}. List of the states in the $\sigma^z$ and $(\phi,\frac{\pi}{2},\alpha)$ bases are listed in the Tables \ref{table:sigma_z_base_L=2} and \ref{table:sigma_x_base_L=2}.
\begin{table}[h]
\centering
\begin{tabular}{|c|c|c|}
\hline
 base configuration& \textbf{$\textbf{R}^{\phi}$} & \textbf{$\Tilde{\textbf{R}}^{\phi}$} \\ 
\hline
$(00)$& $ \begin{pmatrix} 0 & e^{2i\phi}r_{12} \\ -e^{2i\phi}r_{12} & 0 \end{pmatrix}$ & 
$ \frac{1}{a}\begin{pmatrix} 0 & b^{\phi}_{(00)} \\ -b^{\phi}_{(00)} & 0 \end{pmatrix}$ \\
\hline
$(01)$& $ \begin{pmatrix} 0 & r_{12} \\ -r_{12} & 0 \end{pmatrix}$ & 
$ \frac{1}{a}\begin{pmatrix} 0 & b^{\phi}_{(01)} \\ -b^{\phi}_{(01)} & 0 \end{pmatrix}$ \\
\hline
$(10)$& $ \begin{pmatrix} 0 & r_{12} \\ -r_{12} & 0 \end{pmatrix}$ & 
$ \frac{1}{a}\begin{pmatrix} 0 & b^{\phi}_{(10)} \\ -b^{\phi}_{(10)} & 0 \end{pmatrix}$ \\
\hline
$(11)$& $ \begin{pmatrix} 0 & e^{-2i\phi}r_{12} \\ -e^{-2i\phi}r_{12} & 0 \end{pmatrix}$ & 
$ \frac{1}{a}\begin{pmatrix} 0 & b^{\phi}_{(11)} \\ -b^{\phi}_{(11)} & 0 \end{pmatrix}$ \\
\hline
\end{tabular}
\caption{$\textbf{R}^{\phi}$ and  $\Tilde{\textbf{R}}^{\phi}$ for different base configurations for $L=2$.}
\label{tab:matrix_comparison}
\end{table}

\begin{table*}[ht!]
    \centering
    \begin{tabular}{ |m{2.8cm}||m{2.5cm}|m{2.5cm}|m{2.5cm}|m{2.5cm}| }
        \hline
        \multicolumn{5}{|c|}{$L = 2$, $\sigma^z$ basis} \\
        \hline
        \diagbox[innerwidth=2.8cm]{ state}{configs} & \centering $\ket{11}$ & \centering $\ket{10}$ & \centering $\ket{01}$ & \centering $\ket{00}$ \tabularnewline
        \hline
        \centering $\ket{\textbf{R}, 00}$ & \centering $r_{12}$ & \centering $0$ & \centering $0$ & \centering $1$ \tabularnewline
        \hline
        \centering $\ket{\textbf{R},01}$ & \centering $0$ & \centering $r_{12}$ & \centering $1$ & \centering $0$ \tabularnewline
        \hline
        \centering $\ket{\textbf{R},10}$ & \centering $0$ & \centering $1$ & \centering -$r_{12}$ & \centering $0$ \tabularnewline
        \hline
        \centering $\ket{\textbf{R},11}$ & \centering $1$ & \centering $0$ & \centering $0$ & \centering -$r_{12}$ \tabularnewline
        \hline
    \end{tabular}
    \caption{The state in the $\sigma^z$ basis, with a normalization prefactor of $\frac{1}{\mathcal{N}_{R}}$.}
    \label{table:sigma_z_base_L=2}
\end{table*}

\begin{table*}[ht!]
    \centering
    \begin{tabular}{ |m{2.8cm}||m{2.5cm}|m{2.5cm}|m{2.5cm}|m{2.5cm}| }
        \hline
        \multicolumn{5}{|c|}{$L = 2$, $(\phi,\frac{\pi}{2},\alpha)$ basis} \\
        \hline
        \diagbox[innerwidth=2.8cm]{state}{configs} & \centering $\ket{++}$ & \centering $\ket{+-}$ & \centering $\ket{-+}$ & \centering $\ket{--}$ \tabularnewline
        \hline
        \centering $\ket{\textbf{R},00}$ & \centering $a^{\phi}_{(00)}$ & \centering -$b^{\phi}_{(00)}$ & \centering -$b^{\phi}_{(00)}$ & \centering $a^{\phi}_{(00)}$ \tabularnewline
        \hline
        \centering $\ket{\textbf{R},01}$ & \centering $a^{\phi}_{(01)}$ & \centering $b^{\phi}_{(01)}$ & \centering -$b^{\phi}_{(01)}$ & \centering -$a^{\phi}_{(01)}$ \tabularnewline
        \hline
        \centering $\ket{\textbf{R},10}$ & \centering $b^{\phi}_{(10)}$ & \centering -$a^{\phi}_{(10)}$ & \centering $a^{\phi}_{(10)}$ & \centering -$b^{\phi}_{(10)}$ \tabularnewline
        \hline
        \centering $\ket{\textbf{R},11}$ & \centering $b^{\phi}_{(11)}$ & \centering $a^{\phi}_{(11)}$ & \centering $a^{\phi}_{(11)}$ & \centering $b^{\phi}_{(11)}$ \tabularnewline
        \hline
    \end{tabular}
    \caption{ The state in the $(\phi,\frac{\pi}{2},\alpha)$ basis, with a normalization prefactor of $\frac{1}{\sqrt{2}\mathcal{Z}^{\phi}}$.}
    \label{table:sigma_x_base_L=2}
\end{table*}

\subsection{Gaussian pure states with size $L = 3$}

In this subsection, we write the Gaussian state with $L=3$ in the $\sigma^z$ and $\sigma^x$ bases. We first define:

\begin{eqnarray}{\label{abcd}}
\mathcal{N}_R &=&\sqrt{1 + r_{12}^2 + r_{13}^2 + r_{23}^2},\nonumber\\
a&=&1 + r_{12} + r_{13} + r_{23},\nonumber\\
b&=&1 + r_{12} - r_{13} - r_{23},\nonumber\\
c&=&1 - r_{12} + r_{13} - r_{23},\nonumber\\
d&=&1 - r_{12} - r_{13} + r_{23}.\nonumber\\
\mathcal{Z}&=&2\sqrt{1+r_{12}^2+r_{13}^2+r_{23}^2}.\nonumber
\end{eqnarray}
The corresponding antisymmetric matrices are
\begin{equation}
    \bf{R} = \left(
\begin{array}{ccc}
 0 & \text{$r_{12}$} & \text{$r_{13}$} \\
 -\text{$r_{12}$} & 0 & \text{$r_{23}$} \\
 -\text{$r_{13}$} & -\text{$r_{23}$} & 0 \\
\end{array}
\right),
\end{equation}
and
\begin{equation}
\bf{\Tilde{R}} = \frac{1}{\text{$a$}} \left(
\begin{array}{ccc}
 0 & \text{$c$}&
   \text{$d$} \\
  -\text{$c$}& 0 &
   \text{$b$} \\
 -\text{$d$} &
   -\text{$b$} & 0 \\
\end{array}
\right). 
\end{equation} 
List of all the states in the $\sigma^z$ and $\sigma^x$ bases are in the tables \ref{table:L3sigmaz} and \ref{table:L3sigmax}.

\begin{table*}[ht!]
    \centering
    \begin{tabular}{ |m{2.2cm}||m{1.2cm}|m{1.2cm}|m{1.2cm}|m{1.2cm}|m{1.2cm}|m{1.2cm}|m{1.2cm}|m{1.2cm}| }
        \hline
        \multicolumn{9}{|c|}{$L = 3$, $\sigma^z$ basis} \\
        \hline
        \diagbox[innerwidth=2.2cm]{state}{configs} & \centering $\ket{111}$ & \centering $\ket{110}$ & \centering $\ket{101}$ & \centering $\ket{100}$ & \centering $\ket{011}$ & \centering $\ket{010}$ & \centering $\ket{001}$ & \centering $\ket{000}$ \tabularnewline
        \hline
        \centering $\ket{\textbf{R},000}$ & \centering $0$ & \centering $r_{12}$ & \centering $r_{13}$ & \centering $0$ & \centering $r_{23}$ & \centering $0$ & \centering $0$ & \centering $1$ \tabularnewline
        \hline
        \centering $\ket{\textbf{R},001}$ & \centering $r_{12}$ & \centering $0$ & \centering $0$ & \centering $r_{13}$ & \centering $0$ & \centering $r_{23}$ & \centering $1$ & \centering $0$ \tabularnewline
        \hline
        \centering $\ket{\textbf{R},010}$ & \centering -$r_{13}$ & \centering $0$ & \centering $0$ & \centering $r_{12}$ & \centering $0$ & \centering $1$ & \centering -$r_{23}$ & \centering $0$ \tabularnewline
        \hline
        \centering $\ket{\textbf{R},011}$ & \centering $0$ & \centering -$r_{13}$ & \centering $r_{12}$ & \centering $0$ & \centering $1$ & \centering $0$ & \centering $0$ & \centering -$r_{23}$ \tabularnewline
        \hline
        \centering $\ket{\textbf{R},100}$ & \centering $r_{23}$ & \centering $0$ & \centering $0$ & \centering $1$ & \centering $0$ & \centering -$r_{12}$ & \centering -$r_{13}$ & \centering $0$ \tabularnewline
        \hline
        \centering $\ket{\textbf{R},101}$ & \centering $0$ & \centering $r_{23}$ & \centering $1$ & \centering $0$ & \centering -$r_{12}$ & \centering $0$ & \centering $0$ & \centering -$r_{13}$ \tabularnewline
        \hline
        \centering $\ket{\textbf{R},110}$ & \centering $0$ & \centering $1$ & \centering -$r_{23}$ & \centering $0$ & \centering $r_{13}$ & \centering $0$ & \centering $0$ & \centering -$r_{12}$ \tabularnewline
        \hline
        \centering $\ket{\textbf{R},111}$ & \centering $1$ & \centering $0$ & \centering $0$ & \centering -$r_{23}$ & \centering $0$ & \centering $r_{13}$ & \centering -$r_{12}$ & \centering $0$ \tabularnewline
        \hline
    \end{tabular}
    \caption{The state in the $\sigma^z$ basis, with a normalization prefactor of $\frac{1}{\mathcal{N}_{R}}$.}
    \label{table:L3sigmaz}
\end{table*}

\begin{table*}[ht!]
    \centering
    \begin{tabular}{ |m{2.2cm}||m{1.2cm}|m{1.2cm}|m{1.2cm}|m{1.2cm}|m{1.2cm}|m{1.2cm}|m{1.2cm}|m{1.2cm}| }
        \hline
        \multicolumn{9}{|c|}{$L = 3$, $\sigma^x$ basis} \\
        \hline
        \diagbox[innerwidth=2.2cm]{state}{configs} & \centering $\ket{+++}$ & \centering $\ket{++-}$ & \centering $\ket{+-+}$ & \centering $\ket{+--}$ & \centering $\ket{-++}$ & \centering $\ket{-+-}$ & \centering $\ket{--+}$ & \centering $\ket{---}$ \tabularnewline
        \hline
        \centering $\ket{\textbf{R},000}$ & \centering $a$ & \centering -$b$ & \centering -$c$ & \centering $d$ & \centering -$d$ & \centering $c$ & \centering $b$ & \centering -$a$ \tabularnewline
        \hline
        \centering $\ket{\textbf{R},001}$ & \centering $a$ & \centering $b$ & \centering -$c$ & \centering -$d$ & \centering -$d$ & \centering -$c$ & \centering $b$ & \centering $a$ \tabularnewline
        \hline
        \centering $\ket{\textbf{R},010}$ & \centering $b$ & \centering -$a$ & \centering $d$ & \centering -$c$ & \centering -$c$ & \centering $d$ & \centering -$a$ & \centering $b$ \tabularnewline
        \hline
        \centering $\ket{\textbf{R},011}$ & \centering $b$ & \centering $a$ & \centering $d$ & \centering $c$ & \centering -$c$ & \centering -$d$ & \centering -$a$ & \centering -$b$ \tabularnewline
        \hline
        \centering $\ket{\textbf{R},100}$ & \centering $d$ & \centering -$c$ & \centering -$b$ & \centering $a$ & \centering $a$ & \centering -$b$ & \centering -$c$ & \centering $d$ \tabularnewline
        \hline
        \centering $\ket{\textbf{R},101}$ & \centering $d$ & \centering $c$ & \centering -$b$ & \centering -$a$ & \centering $a$ & \centering $b$ & \centering -$c$ & \centering -$d$ \tabularnewline
        \hline
        \centering $\ket{\textbf{R},110}$ & \centering $c$ & \centering -$d$ & \centering $a$ & \centering -$b$ & \centering $b$ & \centering -$a$ & \centering $d$ & \centering -$c$ \tabularnewline
        \hline
        \centering $\ket{\textbf{R},111}$ & \centering $c$ & \centering $d$ & \centering $a$ & \centering $b$ & \centering $b$ & \centering $a$ & \centering $d$ & \centering $c$ \tabularnewline
        \hline
    \end{tabular}
    \caption{The state in the $\sigma^x$ basis, with a normalization prefactor of $\frac{1}{\sqrt{2}\mathcal{Z}}$.}
    \label{table:L3sigmax}
\end{table*}
\newpage
\newpage

\section{Details related to the equation (\ref{R' matrix elements})}\label{sec:AppExamplesMixed}

In this Appendix, we list the exact form of the states $  \ket{\bold{R},101000}$ and $\ket{\bold{R'},011101}$ which is used to get the equation (\ref{R' matrix elements}). We have

\setcounter{equation}{0}

\begin{eqnarray}{\label{|R, 101000}}
    \ket{\bold{R},101000}&=&\frac{1}{\mathcal{N}_R}(\ket{101000}+\nonumber\\    
     &-&r_{12}\ket{011000} -r_{13}\ket{000000} +r_{14}\ket{001100}+r_{15} \ket{001010} +r_{16} \ket{001001} +r_{23}\ket{110000} \nonumber\\
     &-&r_{24}\ket{111100} -r_{25}\ket{111010} -r_{26}\ket{111001}-r_{34} \ket{100100} -r_{35} \ket{100010} -r_{36}\ket{100001} \nonumber\\
     &+&r_{45}\ket{101110} +r_{46}\ket{101101} +r_{56}\ket{101011}+\bold{pf}\ R_{1234}\ket{010100} + \bold{pf}\ R_{1235}\ket{010010} \nonumber\\ 
     &+&\bold{pf}\ R_{1236}\ket{010001}-\bold{pf}\ R_{1245} \ket{011110}-\bold{pf}\ R_{1246} \ket{011101} - \bold{pf}\ R_{1256}\ket{011011} \nonumber\\
     &-&\bold{pf}\ R_{1345}\ket{000110} - \bold{pf}\ R_{1346}\ket{000101}-\bold{pf}\ R_{1356}\ket{000011}+\bold{pf}\ R_{1456} \ket{001111} \nonumber\\
     &+&\bold{pf}\ R_{2345} \ket{110110} +\bold{pf}\ R_{2346}\ket{110101}+\bold{pf}\ R_{2356}\ket{110011} - \bold{pf}\ R_{2456}\ket{111111} \nonumber\\
     &-&\bold{pf}\ R_{3456}\ket{100111}+\bold{pf}\ R \ket{010111} )\nonumber
\end{eqnarray}
and 
\begin{eqnarray}{\label{|R', 011101}}
    \ket{\bold{R'},011101}&=&\frac{1}{\mathcal{N}_R'}(\ket{011101}+\nonumber\\    
     &+&r'_{12}\ket{101101} -r'_{13}\ket{110101} +r'_{14}\ket{111001}-r'_{15} \ket{111111} -r'_{16} \ket{111100} -r'_{23}\ket{000101} \nonumber\\
     &+&r'_{24}\ket{001001} -r'_{25}\ket{001111} -r'_{26}\ket{001100}-r'_{34} \ket{010001} +r'_{35} \ket{010111} +r'_{36}\ket{010100} \nonumber\\
     &-&r'_{45}\ket{011011} -r'_{46}\ket{011000} +r'_{56}\ket{011110}-\bold{pf}\ R'_{1234}\ket{100001} + \bold{pf}\ R'_{1235}\ket{100111} \nonumber\\ 
     &+&\bold{pf}\ R'_{1236}\ket{100100}-\bold{pf}\ R'_{1245} \ket{101011}-\bold{pf}\ R'_{1246} \ket{101000} + \bold{pf}\ R'_{1256}\ket{101110} \nonumber\\
     &+&\bold{pf}\ R'_{1345}\ket{110011} + \bold{pf}\ R'_{1346}\ket{110000}-\bold{pf}\ R'_{1356}\ket{110110}+\bold{pf}\ R'_{1456} \ket{111010} \nonumber\\
     &+&\bold{pf}\ R'_{2345} \ket{000011} +\bold{pf}\ R'_{2346}\ket{000000}-\bold{pf}\ R'_{2356}\ket{000110} + \bold{pf}\ R'_{2456}\ket{001010} \nonumber\\
     &-&\bold{pf}\ R'_{3456}\ket{010010}-\bold{pf}\ R' \ket{100010} ).\nonumber
\end{eqnarray}

%\end{appendices}

%\section{References}
%\bibstyle{SciPost_bibstyle}
%\bibliography{Bibliography.bib}

%%%%%%%%%%%%%%%%%%%%%%%%%%%%\end{document}

%\hfill\\
%\newpage
%\hfill\\
%\newpage

\end{document}